\documentclass[prb,twocolumn,showpacs,amsmath,amssymb]{revtex4}
\usepackage{graphicx}
\usepackage{dcolumn}
\usepackage{bm}
\usepackage{color}
\newcommand{\nn}{\nonumber}
\newcommand{\beq}{\begin{eqnarray}}
\newcommand{\eeq}{\end{eqnarray}}

\begin{document}

\title{
Superfluid $^3$He in a restricted geometry with a perpendicular magnetic field
}

\author{Takeshi Mizushima}
\email{mizushima@mp.okayama-u.ac.jp}
\affiliation{Department of Physics, Okayama University,
Okayama 700-8530, Japan}
\date{\today}

\begin{abstract}

We theoretically investigate the role of surface Andreev bound states (SABSs) on the phase diagram and spin susceptibilities of superfluid $^3$He confined to a restricted geometry. We first explicitly derive the dispersion of the SABS in $^3$He-B in the presence of a magnetic field, where the Majorana Ising spin and the spin susceptibility contributed from the SABS are associated with the $SO(3)$ order parameter manifold. Subsequently, based on the quasiclassical Eilenberger theory with Fermi liquid corrections, we discuss the nonlinear effect of a magnetic field on the SABS, where the magnetic field is perpendicular to the specular surface. It is directly demonstrated that a gapped SABS strongly enhances the magnetization density and spin susceptibility at the surface, compared with that in the normal $^3$He. To capture the characteristics of the SABS, we show the field- and temperature-dependences of the spatially averaged susceptibility which is detectable through NMR experiments. It turns out that the contribution of the SABS leads to nonmonotonic temperature-dependence of the spin susceptibility. Furthermore, we present the superfluid phase diagram, where the B-phase undergoes a first-order (second-order) phase transition to A-phase or planar phase at low (high) temperatures.

\end{abstract}



\maketitle

\section{Introduction}

The study of spin-triplet $p$-wave superfluid $^3$He in a restricted geometry has a long history since the mid 1970s. It is known that the most symmetric superfluid phase, called the B-phase, is a ground state in the bulk $^3$He at low temperatures, because the isotropic energy gap gains the more condensation energy than other competitive phases.~\cite{vollhardt,leggettRMP,Wheatley} In a restricted geometry, however, surfaces give rise to the pair breaking, which squashes the isotropic B-phase elliptically. As two surfaces get close to each other, the B-phase at last undergoes a change to the A-phase which has point nodes at the north and south poles of the Fermi sphere. The A-B phase transition induced by a surface boundary condition was directly observed in NMR experiments and a torsional oscillator,~\cite{freemanPRL1988,freemanPRB1990,miyawaki,kawae,kawasaki,levitin,bennett,xu} where superfluid $^3$He is confined to a geometry with submicron thickness. 

Theoretical studies for understanding the pair breaking effect and the $\hat{\bm n}$-texture on surfaces were initiated by the analysis based on the Ginzburg-Landau theory.~\cite{ambegaokar,barton,kjaldman,fujita,takagi,jacobsen,fetter,li,ullah,smith,lin-liu,salomaa,brinkman} Beyond the theory which does not take account of the information on quasiparticles, the quasiclassical Eilenberger theory provides a tractable and quantitative scheme to study the interplay of the pair potential and quasiparticle states.~\cite{qct,serene} It was found by Buchholtz and Zwicknagl~\cite{buchholtzPRB1981} in 1981 that from the microscopic point of view, midgap bound states emerge on a specular surface of the B-phase. The midgap state was more explicitly discussed in Ref.~\onlinecite{haraPTP1986} which analyzed a $p$-wave polar state as an exactly soluble model. Based on the quasiclassical theory, the quantitative phase diagram and the finite size effect of the superfluid $^3$He in a restricted geometry have been clarified in Refs.~\onlinecite{haraJLTP1988,vorontsov,tsutsumi,tsutsumiPRB}, which underline the role of the SABS on thermodynamics. Note that these previous works have not taken account of the effect of a Zeeman magnetic field. Several experiments have observed the pair breaking effect and the enhancement of the surface density of states due to the midgap state,~\cite{osheroff,ahonen,ishikawa,castelijns,aoki,choi,saitoh,wada} concurrently with theoretical works. It is worth mentioning that at the present time, the midgap state is recognized as a family of the Andreev bound state, called the surface Andreev bound state (SABS),~\cite{nagai}. Since this midgap bound state exists only if the superconducting pair potential changes its sign,~\cite{ohashi} it ubiquitously appears in various physics systems, such as superconducting junctions,~\cite{kashiwaya} superconducting vortices,~\cite{CdGM,kaneko} unconventional superconductors,~\cite{hu,tanaka,TM2008} and Fulde-Ferrell-Larkin-Ovchinnikov superconductors.~\cite{machida,TMPRL2005} In addition, the same physics is shared with the solitons which emerge in polyacetylene,\cite{takayama,brazouskii,nakahara,horovitz} the incommensurate spin-density wave,~\cite{machida-fujita,fujita-machida}  and the stripe state in high $T_{\rm c}$ cuprates.~\cite{machida1989} 

Recently, the study of the SABS in superfluid $^3$He-B has been rekindled by the understanding of the direct correspondence to a bulk topological invariant. The chiral symmetry constructed from the time-reversal and particle-hole operations allows us to introduce the three-dimensional winding number which behaves as a topological invariant in $^3$He-B.~\cite{review1,review2,schnyder,qi,volovik2009,ryu,sato09,sato10,sato2009,sato2011,kitaev} Hence, the gapless SABS in $^3$He-B is protected by the topological invariant associated with the time-reversal symmetry. Furthermore, the chiral symmetry ensures that the SABS behaves as a Majorana fermion which is a particle equivalent to its own anti-particle.~\cite{chung,sato2011,TM2012} As a consequence of a Majorana fermion and the topological protected bound state, it is unveiled that the multifaceted SABS is insensitive to the density fluctuation and exhibits Ising-like isotropy of magnetic response, that is, what is known as the Majorana Ising spin.~\cite{chung,nagato,shindou} Recently, the gapless cone spectrum has been observed by measuring the transverse acoustic impedance on surface of $^3$He-B with controlled specularity.~\cite{murakawaPRL2009,murakawaJPSJ2011}

The Majorana fermion and Ising spin may survive even when a magnetic field is applied along a direction parallel to the specular surface, where the time-reversal symmetry is broken. In a weak magnetic field comparable with the dipolar field, the B-phase stays topological as a consequence of the protection with a hidden ${\bm Z}_2$ symmetry, that is, called the symmetry protected topological order.~\cite{TM2012} This ensures the chiral symmetry and protects the gapless Majorana cone even in the presence of a magnetic field. In the case of a perpendicular field, however, an infinitesimal magnetic field is always able to open a finite energy gap in the surface Majorana cone. Within the linear regime of the applied field, the spin susceptibility with respect to a perpendicular magnetic field is considerably enhanced, compared with that in the normal $^3$He.~\cite{nagato} This enhancement is expected to make the SABS detectable in NMR experiments. Nevertheless, there have never been any studies which quantitatively and microscopically discuss thermodynamics and spin susceptibility in the nonlinear regime of the magnetic field. Note that in the magnetic field regime much stronger than the dipolar field ($\sim \! 30{\rm G}$), the Majorana Ising anisotropy disappears and the thermodynamics is independent of the orientation of the applied magnetic field.~\cite{volovik2010,silaev,TM2012}

\begin{figure}[b!]
\includegraphics[width=65mm]{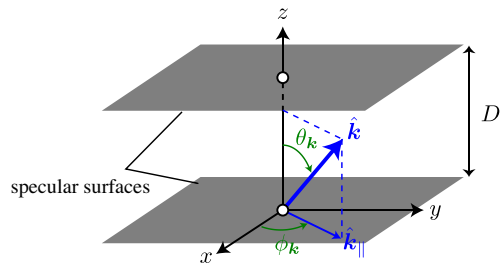}
\caption{(Color online) Schematic picture of the system which is considered in this work. The superfluid $^3$He is sandwiched by two specular walls whose distance is set to be $D$. In and after Sec.~III, a magnetic field is applied to the $z$-axis which is normal to the wall. }
\label{fig:system}
\end{figure}

Hence, the purpose of this paper is to clarify the role of the SABS on thermodynamics and spin susceptibility in superfluid $^3$He restricted to a slab geometry. As displayed in Fig.~\ref{fig:system}, we set $^3$He to be sandwiched by two specular walls and a magnetic field is applied along the $z$-axis which is normal to the wall. First, we clarify that the Majorana nature of the SABS is associated with the $SO(3)$ order parameter manifold of $^3$He-B. Based on the quasiclassical Eilenberger theory taking account of Fermi liquid corrections, we quantitatively discuss the thermodynamics and the enhancement of the magnetization density due to a gapped SABS in a slab geometry. We here present the quantitative superfluid phase diagram in $^3$He restricted to a slab, where it is emphasized that the Fermi liquid correction plays a critical role on determining the A-B phase boundary induced by the magnetic field. To capture the characteristics in experiments, we clarify the field- and temperature-dependences of the spatially averaged spin susceptibility, which is detectable in NMR experiments. In particular, we emphasize the nonlinear effect of the applied magnetic field. It turns out that the spatially averaged spin susceptibility in a $^3$He with a submicron thickness exhibits the non-monotonic behavior on the temperature. 

In the following section, we explicitly derive the dispersion of the surface bound states in the B-phase parameterized with the $\hat{\bm n}$-vector and $\varphi$ which are the parameter of the $SO(3)$ order parameter manifold. Here, we discuss the relation between the $SO(3)$ manifold and spin susceptibility contributed from the surface bound state. Then, we move on to the quantitative calculation based on the quasiclassical Eilenberger theory. In Sec.~III, we describe the details of the quasiclassical formulation for superfluid $^3$He in a restricted geometry, where the self-consistent framework takes account of the Fermi liquid corrections. The self-consistent solutions for the pair potential, magnetization density, and local density of states are presented in Sec.~IV, where the contribution of the surface bound state to magnetization density is underlined. The complete phase diagram is proposed in Sec.~V, where we demonstrate that the phase boundary is sensitive to the Fermi liquid corrections. We also present the non-linear effect of the Zeeman magnetic field on the spin susceptibility. The final section is devoted to conclusion and discussion.  The details on the derivation of the dispersion of the SABS are described in Appendix A. The numerical procedure and boundary condition are given in Appendix B and Appendix C shows that the quasiclassical Eilenberger equation for $^3$He-B is invariant under an $SO(2)$ rotation in spin and orbital spaces, which shortens computation time. Throughout this paper, we set $\hbar\!=\! k_{\rm B}\!=\! 1$ and repeated Greek (Roman) indices imply the sum over $x,y,z$ ($\uparrow,\downarrow$).

\section{Surface bound state and Majorana Ising spin in the B-phase}

\subsection{The B-phase order parameter}

Let us start with the mean-field Hamiltonian density for superfluid $^3$He with the mass $M$ in the $4\!\times\! 4$ Nambu representation,
\beq
\underline{\mathcal{H}}({\bm r}_1,{\bm r}_2) = \delta ({\bm r}_{12})
\left[ \epsilon ({\bm r}_1)\underline{\tau}_z + \underline{V} 
\right]  
+ \underline{\Sigma} ({\bm r}_1,{\bm r}_2) .
\label{eq:hami}
\eeq
The single-particle Hamiltonian density is $\epsilon ({\bm r})\!=\!-{\bm \nabla}^2/2M-E_{\rm F}$ and the Zeeman energy $\underline{V}$ is given by
\beq
\underline{V} \equiv -\mu _{\rm n}H_{\mu}\left(
\begin{array}{cc}
\sigma _{\mu} & 0 \\ 
0 & -\sigma^{\ast}_{\mu} \end{array}\right), 
\eeq
with the Fermi energy $E_{\rm F}\!=\!k^2_{\rm F}/2M$, and magnetic moment of $^3$He nuclei $\mu _{\rm n}$. We also introduce the Pauli matrices $\sigma _{\mu}$ ($\tau _{\mu}$) in the spin (particle-hole) spaces. The self-energy matrix $\underline{\Sigma}({\bm r}_1,{\bm r}_2)$ consists of the Fermi liquid correction $\underline{\Sigma}_{\rm FL}$ and the pair potential $\underline{\Delta}$, that is, $\underline{\Sigma}\!\equiv\! \underline{\Sigma}_{\rm FL} + \underline{\Delta}$, and 
\beq
\underline{\Delta}({\bm r}_1,{\bm r}_2) \equiv 
\left[
\begin{array}{cc}
0 & \Delta ({\bm r}_1,{\bm r}_2)  \\
-\Delta^{\ast}({\bm r}_1,{\bm r}_2) &  0 \end{array}\right],
\eeq
where $\Delta $ is a $2\!\times\! 2$ matrix in the spin space and the spin triplet and $p$-wave symmetries require $\Delta _{ab}({\bm r}_1,{\bm r}_2) \!=\! \Delta _{ba}({\bm r}_1,{\bm r}_2)$ and $\Delta _{ab}({\bm r}_1,{\bm r}_2) \!=\! -\Delta _{ab}({\bm r}_2,{\bm r}_1)$.

Without loss of generality, a pair potential for a spin-triplet superfluid is expressed with the ${\bm d}$-vector as $\Delta ({\bm k},{\bm r}) \!\equiv\! \int d{\bm r}_{12}e^{-i{\bm k}\cdot{\bm r}_{12}} \Delta ({\bm r}_1,{\bm r}_2)\!=\! i\sigma _{\mu}\sigma _y d_{\mu}(\hat{\bm k},{\bm r})$. The superfluid $^3$He-B phase is known to be most symmetric among possible order parameters, which is invariant under the joint rotations of three-dimensional spin and orbital spaces, $SO(3)_{{\bm L}+{\bm S}}$. The order parameter is the eigenstate of the angular momentum operator composed of the spin and orbital angular momentum, ${\bm S}$ and ${\bm L}$,
\beq
J_{\mu} = L_{\mu} + S_{\nu}R_{\nu\mu}(\hat{\bm n},\varphi),
\eeq
implying the spontaneously broken spin-orbit symmetry, $SO(3)_{{\bm L}-{\bm S}}$, in addition to the ordinary $U(1)$ gauge $\vartheta$.~\cite{vollhardt,leggettRMP} Here, $R_{\mu\nu}(\hat{\bm n},\varphi)$ describes the relative rotation matrix between spin and orbital spaces originated from the $SO(3)_{{\bm L}-{\bm S}}$ manifold, where $\hat{\bm n}$ and $\varphi$ denote the rotation axis and angle. Then, the general form of the order parameter of the superfluid $^3$He-B is described as $d_{\mu}(\hat{\bm k},{\bm r}) \!=\! d_{\mu\nu}({\bm r})\hat{k}_{\nu}$
\beq
d_{\mu \nu}({\bm r}) = e^{i\vartheta}R_{\mu \nu} (\hat{\bm n},\varphi) \Delta _{\nu}({\bm r}).
\label{eq:op}
\eeq
This order parameter also describes the planar phase with $\Delta _z \!=\! 0$ and squashed B (or B$_2$) phase with $\Delta _z \!\neq\! \Delta _x \!=\! \Delta _y$.~\cite{vollhardt,leggettRMP} Since the planar phase is energetically degenerate with the A-phase at the weak coupling limit, the order parameter in Eq.~(\ref{eq:op}) takes account of all possible phases stabilized in the presence of a magnetic field.

Here, to derive the SABS from the general form in Eq.~(\ref{eq:op}), let us consider the situation where a single specular surface is set at $z\!=\! 0$. The energy eigenstates of the Hamiltonian in Eq.~(\ref{eq:hami}) are obtained by solving the following eigenvalue equation, 
\beq
\int d{\bm r}_2 \mathcal{H}({\bm r}_1,{\bm r}_2) {\bm \varphi}({\bm r}_2) = E {\bm \varphi}({\bm r}_2),
\label{eq:bdg}
\eeq
which is called the Bogoliubov-de Gennes (BdG) equation. Here, the wavefunctions obey the normalization condition, $\int d{\bm r}{\bm \varphi}^{\dag}({\bm r}){\bm \varphi}({\bm r}) \!=\! 1$. To solve Eq.~(\ref{eq:bdg}) analytically, in this section, we ignore the Fermi liquid correction, that is, $\underline{\Sigma}_{\rm FL} \!=\! 0$. The effect is discussed in the subsequent sections with the quasiclassical Eilenberger theory. In addition, the pair potential $\Delta$ is assumed to be spatially uniform $\Delta (\hat{\bm k},{\bm r}) \!=\! \Delta (\hat{\bm k})$. It is often convenient to utilize the alternative description with $U(\hat{\bm n},\varphi) \!\in\! {\rm SU}(2)$ as $U(\hat{\bm n},\varphi)\sigma _{\nu}U^{\dag}(\hat{\bm n},\varphi) \!=\! \sigma _{\mu}R_{\mu\nu}(\hat{\bm n},\varphi)$. Using this ${\rm SU}(2)$ matrix, the B-phase order parameter in Eq.~(\ref{eq:op}) reduces to
\beq
{\Delta}(\hat{\bm k},{\bm r}) = U(\hat{\bm n},\varphi) {\Delta}_0(\hat{\bm k},{\bm r}) U^{\rm T}(\hat{\bm n},\varphi),
\label{eq:opsu2}
\eeq
where $U^{\rm T}$ denotes the transpose of a matrix $U$ and ${\Delta}_0(\hat{\bm k},{\bm r}) \!=\! i e^{i\vartheta}\sigma _{\mu}\sigma _y \Delta_{\mu}({\bm r})\hat{k}_{\mu}$ is the simplest expression of the B-phase order parameter.

Using the Andreev approximation which holds within the weak coupling regime $k_{\rm F}\xi \!=\! 2E_{\rm F}/\Delta \!\gg\! 1$, the BdG equation (\ref{eq:bdg}) reduces to the Andreev equation 
\beq
\left[ - i \alpha v_{\rm F}\cos\theta _{\bm k}\partial _z\underline{\tau}_z + \underline{V}
+ \underline{\Delta} ({\bm k}_{{\rm F},\alpha})\right] \tilde{\bm \varphi}_{\alpha}(z) = E\tilde{\bm \varphi}_{\alpha}(z),
\label{eq:andr}
\eeq
where $\tilde{\bm \varphi}_{\pm} (z)$ describes the slowly varying part of quasiparticle wavefunction ${\bm \varphi}({\bm r})$, that is, ${\bm \varphi}({\bm r}) \!=\! \sum _{\alpha \!=\! \pm} C_{\alpha}\tilde{\bm \varphi}_{\alpha}(z)e^{i{\bm k}_{{\rm F},\alpha}\cdot{\bm r}}$ with the Fermi velocity $v_{\rm F}$ and ${\bm k}_{{\rm F},\alpha} \!=\! k_{\rm F}(\cos\!\phi _{\bm k}\sin\!\theta _{\bm k},\sin\!\phi _{\bm k}\sin\!\theta _{\bm k},\alpha\cos _{\bm k}\!\theta _{\bm k})$.

\subsection{Majorana Ising spin}

First, we solve the Andreev equation (\ref{eq:andr}) in the absence of a magnetic field, $H\!=\! 0$, where the resulting equation becomes equivalent to that for spinless chiral $p$-wave superconductors.~\cite{stone} As described in Appendix A, the bound state solution with $|E({\bm k}_{\parallel})| \!\le\! \Delta _0$ has the energy dispersion linear on the momentum ${\bm k}_{\parallel} \!=\! (k_x,k_y)$ as
\beq
E_0({\bm k}_{\parallel}) = \pm \frac{\Delta _0}{k_{\rm F}} |{\bm k}_{\parallel}|.
\label{eq:E0a}
\eeq 
This expression is independent of the orientation of $\hat{\bm n}$ and the angle $\varphi$. The corresponding wavefunctions for the quasiparticles bound at at $z\!=\! 0$ are given by
\beq
{\bm \varphi}^{(\pm)}_{0,{\bm k}_{\parallel}} ({\bm r}) = N_{\bm k}
e^{i{\bm k}_{\parallel}\cdot{\bm r}_{\parallel}}f(k_{\perp},z)\mathcal{U}(\hat{\bm n},\varphi)
{\bm \Phi}_{\pm}(\phi _{\bm k}), 
\label{eq:varphi1}
\eeq
where $N_{\bm k}$ is the normalization constant and $\mathcal{U} \!\equiv\! {\rm diag}(U,U^{\ast})$. In Eq.~(\ref{eq:varphi1}), we also set $f(k_{\perp},z)\!=\! \sin\left( k_{\perp}z\right)e^{-z/\xi}$ with $k_{\perp} \!\equiv\! \sqrt{k^2_{\rm F}-k^2_{\parallel}}$ and 
\beq
{\bm \Phi}_{\pm}(\phi _{\bm k}) \equiv e^{\pm i\frac{\phi _{\bm k}}{2}}\left[
e^{-i\frac{\phi _{\bm k}}{2}}
\left( 
\begin{array}{c}
1 \\ 0 \\ 0 \\ -i
\end{array}
\right)
\mp e^{i\frac{\phi _{\bm k}}{2}}\left( 
\begin{array}{c}
0 \\ i \\ 1 \\ 0
\end{array}
\right) \right].
\eeq
The wavefunction ${\bm \varphi}^{(+)}_{0,{\bm k}_{\parallel}}$ corresponds to the positive energy solution of $E_0({\bm k}_{\parallel})$ and ${\bm \varphi}^{(-)}_{0,{\bm k}_{\parallel}}$ is the negative branch. The particle-hole symmetry $\underline{\tau}_x \underline{\mathcal{H}}({\bm k},{\bm r})\underline{\tau}^{\dag}_x \!=\! -\mathcal{H}^{\ast}(-{\bm k},{\bm r})$ ensures the one-to-one correspondence between the two branches as ${\bm \varphi}^{(-)}_{0,{\bm k}_{\parallel}} \!=\! \underline{\tau}_x {\bm \varphi}^{(+)\ast}_{0,-{\bm k}_{\parallel}}$.

The quantized field ${\bm \Psi}\!=\! (\Psi _{\uparrow}, \Psi _{\downarrow},\Psi^{\dag}_{\uparrow}, \Psi^{\dag}_{\downarrow})^{\rm T}$ in spin-triplet superfluids can be expanded in terms of the positive energy states of the SABS with $E({\bm k}_{\parallel}) \!\ge\!0$ and ${\bm \varphi}_{{\bm k}_{\parallel}}({\bm r})$ in addition to continuum states. For low temperature regimes $T\!\ll\!\Delta _0$, the field operator can be constructed from the contributions of only the SABS as ${\bm \Psi}({\bm r}) \!\approx\! \sum _{{\bm k}_{\parallel}} [ {\bm \varphi}^{(+)}_{0,{\bm k}_{\parallel}}({\bm r})\eta _{{\bm k}_{\parallel}}+\underline{\tau}_x {\bm \varphi}^{(+)\ast}_{0,{\bm k}_{\parallel}}({\bm r})\eta^{\dag}_{{\bm k}_{\parallel}}]$, where $\eta _{{\bm k}_{\parallel}}$ and $\eta^{\dag}_{{\bm k}_{\parallel}}$ denote the Bogoliubov quasiparticle operators. Then, the field operator contributed from the SABS obeys the self-conjugate Majorana condition,
\beq
\Psi^{\rm M}_{a} ({\bm r}) \approx - \left[{\Psi}^{\rm M}_{a}({\bm r})\right]^{\dag},
\label{eq:majorana}
\eeq
where $\Psi^{\rm M}_{a} ({\bm r}) \!\equiv\!
{U}^{\prime\dag}_{ab}(\hat{\bm n},\varphi) {\Psi}^{\prime}_{b}({\bm r})$. Using the wavefunction of the SABS, ${\bm \Psi}^{\prime}_{\beta}({\bm r})$ is given as 
$
{\bm \Psi}^{\prime}({\bm r}) = \sum _{{\bm k}_{\parallel}} [
e^{i{\bm k}_{\parallel}\cdot{\bm r}_{\parallel}+i\phi _{\bm k}/2}\eta _{{\bm k}_{\parallel}} 
-{\rm h.c.}
]\mathcal{U}^{\prime}(\hat{\bm n},\varphi)
{\bm \Phi}^{\prime}_{\bm k}$.
Here, the spin quantization axis is changed from the $\hat{\bm z}$ to $\hat{\bm x}$-axis, where ${\bm \Psi}$, $\mathcal{U}$, ${\bm \Phi}^{(\pm)}_{\bm k}$ change to ${\bm \Psi}^{\prime}$, $\mathcal{U}^{\prime}$, ${\bm \Phi}^{\prime}_{\bm k}\!\equiv\! [\cos\!\frac{\bar{\phi}_{\bm k}}{2},\sin\!\frac{\bar{\phi}_{\bm k}}{2},-\cos\!\frac{\bar{\phi}_{\bm k}}{2},-\sin\!\frac{\bar{\phi}_{\bm k}}{2}]^{\rm T}$.

Once Eq.~(\ref{eq:majorana}) holds, it is straightforward to prove that the Majorana fields ${\Psi}^{\rm M}_{a}$ behave as the Clifford algebra, $\{ \Psi^{\rm M}_{a}({\bm r}_1), \Psi^{\rm M}_{b} ({\bm r}_2)\} \!=\! 2\delta _{a,b} \delta ({\bm r}_{12})$. For the case of $\hat{\bm n}\!\parallel\! \hat{\bm z}$, the local spin operator $S_{\mu}({\bm r})\!\equiv\! \frac{1}{2}\Psi^{\dag}_{a}({\bm r})(\sigma _{\mu})_{ab}\Psi _{b}({\bm r})$ with the Clifford algebra results in the Ising-like anisotropic form as ${\bm S} \!=\! (0,0,S^{\rm M}_z) \!\equiv\! {\bm S}^{\rm M}$, where $S^{\rm M}_{z} \!\equiv\! - \frac{1}{2}\Psi^{\rm M}_a(\sigma _{\mu})_{ab}\Psi^{\rm M}_b$. For an arbitrary configuration of $(\hat{\bm n},\varphi)$, the local spin operator contributed from the SABS results in
\beq
S_{\mu}({\bm r}) \!=\! R_{\mu z}(\hat{\bm n},\varphi)S^{\rm M}_{z}({\bm r}).
\label{eq:chi}
\eeq
The direction of the Majorana Ising spin reflects the $SO(3)$ order parameter manifold $(\hat{\bm n},\varphi)$. Using Eq.~(\ref{eq:chi}), the dynamical spin susceptibility becomes $\chi _{\mu\nu}({\bm r}_1,{\bm r}_2; \omega) \!=\! \chi^{\rm M}_{zz}({\bm r}_1,{\bm r}_2;\omega)
R_{\mu z}(\hat{\bm n},\varphi) R_{\nu z} (\hat{\bm n},\varphi)$. This implies that magnetization and susceptibility originate from Majorana Ising spins ${S}^{\rm M}_{\mu}({\bm r})$ and $\chi^{({\rm M})}_{zz}({\bm r}_1,{\bm r}_2;\omega)\!\equiv\! \langle S^{\rm M}_{z}({\bm r}_1)S^{\rm M}_{z}({\bm r}_2)\rangle _{\omega}$ through the $SO(3)$ matrix $R_{\mu\nu}(\hat{\bm n},\varphi)$. The property of $\chi^{({\rm M})}_{zz}({\bm r}_1,{\bm r}_2;\omega)$ was discussed in Ref.~\onlinecite{chung,TM}.

\begin{figure}[b!]
\includegraphics[width=50mm]{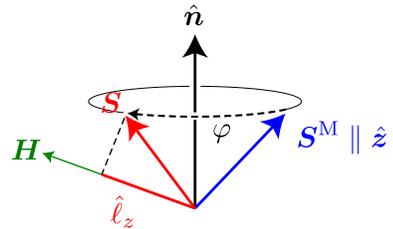}
\caption{(Color online) Schematic picture on the relation between $\hat{\ell}_z$ and ${\bm S}$, where ${\bm S}$ and ${\bm S}^{\rm M}$ denote the orientation of the Majorana Ising spins for an arbitrary $\hat{\bm n}$ and ${\bm S}$ for $\hat{\bm n} \!\parallel\! \hat{\bm z}$. The $\hat{\bm z}$-axis is normal to the specular surfaces as displayed in Fig.~\ref{fig:system}.}
\label{fig:lz_pix}
\end{figure}

To understand the orientation of the Majorana Ising spin in Eq.~(\ref{eq:chi}), it is convenient to introduce the $\hat{\bm \ell}$-vector in Eq.~(\ref{eq:sabs}), the definition~\cite{volovik2010,TM2012} of which is
\beq
\hat{\ell}_{\mu} (\hat{\bm n}, \varphi) \equiv \hat{h}_{\nu} R_{\nu \mu} (\hat{\bm n}, \varphi).
\eeq
The orientation of an applied magnetic field is denoted by $\hat{h}_{\nu} \!=\! H_{\nu}/H$. Then, it turns out that the $\hat{\ell}_z(\hat{\bm n}, \varphi)$ describes the projection of the Majorana Ising spin ${\bm S}({\bm r})$ in Eq.~(\ref{eq:chi}) onto the orientation of the applied magnetic field ${\bm H}$ as
\beq
\hat{\ell}_z(\hat{\bm n},\varphi) = \frac{\hat{\bm h}\cdot{\bm S}({\bm r})}{\left| {\bm S}({\bm r}) \right|}.
\label{eq:lz}
\eeq
Figure~\ref{fig:lz_pix} depicts the schematic picture for ${\bm S}$, ${\bm H}$, and $\hat{\ell}_z$. For $\hat{\ell}_z \!=\! 0$, the Majorana Ising spin ${\bm S}$ is perpendicular to the applied magnetic field, which implies that the SABS does not contribute to the magnetic response. However, the SABS may be responsible to ${\bm H}$ when $\hat{\ell}_z\!\neq\! 0$.  

The gapless spectrum of the SABS is protected by the nontrivial topological invariant defined in the bulk region of the B-phase.~\cite{schnyder,qi,TM2012} As two specular surfaces at $z \!=\! 0$ and $D$ get close to each other, however, the interference between the SABSs distorts the surface cone spectrum in Eq.~(\ref{eq:E0a}), where $D$ denotes the thickness of the sample. Then, the hybridization of the two SABSs exponentially splits the zero energy state at $|{\bm k}_{\parallel} |\!=\! 0$ with quantum oscillation on the scale of $k^{-1}_{\rm F}$ as $\delta E({\bm k}_{\parallel} \!=\! {\bm 0}) \!\sim\! e^{-D/\xi}\sin(k_{\rm F}D)$.~\cite{TM2010v2,kawakami,cheng1,cheng2} In the quasiclassical Eilenberger theory, the quantum oscillation term vanishes.~\cite{tsutsumiPRB} In addition to the splitting due to the quasiparticle tunneling, the finite size of the system with the thickness $D \!=\! \mathcal{O}(\xi)$ gives rise to the pair breaking effect, which may also stimulate the deviation of the gapless spectrum. The distortion of the gapless Majorana cone due to the quasiparticle tunneling and pair breaking effect may break the Majorana Ising nature of the surface bound states. The numerical analysis on this issue will be discussed in Sec.~IV.

Now let us turn to the case of a finite magnetic field $H\!\neq\! 0$. As described in Appendix A, the dispersion of the SABS is given as
\beq
E({\bm k}_{\parallel}) = \pm \sqrt{ \left| E_0({\bm k}_{\parallel})\right|^2 + \left|\mu _{\rm n}H\hat{\ell}_z(\hat{\bm n}, \varphi)\right|^2},
\label{eq:sabs}
\eeq
and the wave functions are obtained from Eq.~(\ref{eq:eigenfn}) with $a_{\pm}({\bm k}_{\parallel}) \!=\! \sqrt{\frac{1}{2}(1\pm |\frac{E_0({\bm k}_{\parallel})}{E({\bm k}_{\parallel})}|)}$. The resulting dispersion in Eq.~(\ref{eq:sabs}) implies that the energy gap of the surface state depends on the $\hat{\bm \ell}$-vector as
\beq
\min \left|E ({\bm k}_{\parallel})\right| = \mu _{\rm n}H  \left|\hat{\ell}_z(\hat{\bm n},\varphi)\right|.
\label{eq:minE}
\eeq
For $\hat{\bm n}\!=\! \hat{\bm z}$, since $\hat{\ell}_z \!=\! \hat{h}_z$, the dispersion in Eq.~(\ref{eq:sabs}) is consistent with the previous works in Refs.~\onlinecite{chung,nagato,volovik2009,shindou}. Equation~(\ref{eq:sabs}) indicates that for $\hat{\ell}_z \!\neq\! 0$, the Majorana Ising nature of the SABS disappears and an arbitrary orientation of the magnetic field opens a finite energy gap.~\cite{volovik2010} The Ising anisotropy is also consistent with Eq.~(\ref{eq:lz}) describing the relation between the orientation of the Majorana Ising spin and the applied field.

\subsection{Effect of a Zeeman magnetic field and the dipole interaction}

It is important to mention the relation between the energy gap of the SABS in Eq.~(\ref{eq:minE}) and the stable configuration of $(\hat{\bm n}, \varphi)$. The magnetic field energy density within the Ginzburg-Landau theory is given as
\beq
F_{\rm H} = - \epsilon _{\rm S}  \left[H_{\mu}R_{\mu \nu}(\hat{\bm n},\varphi)\hat{s}_{\nu}\right]^2 
= - \epsilon _{\rm S} H^2\left| \hat{\ell}_z (\hat{\bm n},\varphi)\right|^2,
\eeq
where $\epsilon _{\rm S} \!\equiv \! \xi _0 (\chi _N-\chi _B) \!>\! 0$.~\cite{brinkman} The minimization condition of $F_{\rm H}$, that is, $|\hat{\ell}_{z}(\hat{\bm n},\varphi) |\!=\! 1$, opens the maximum energy gap of the surface Andreev bound state in Eq.~(\ref{eq:minE}), $\min |E({\bm k}_{\parallel})| \!=\! \mu _{\rm n}H$. 

On the other hand, the dipole energy favors to align $\hat{\bm n}$ to the $\hat{\bm z}$-axis. Hence, it turns out that a Zeeman magnetic field perpendicular to the surface always open a finite energy gap in the SABS, $\min |E({\bm k}_{\parallel})| \!=\! \mu _{\rm n}H$, because both the magnetic field energy and dipole energy favor $\hat{\bm n}\!\parallel\! \hat{\bm z}$. In the case of a parallel magnetic field, however, the dipole interaction energy may be competitive to the magnetic field energy. Actually, it is demonstrated in Ref.~\onlinecite{TM2012} that the gapless SABS with $\hat{\ell}_z$ may be protected by a hidden ${\bm Z}_2$ symmetry which is preserved under a magnetic field weaker than the dipolar field is regarded as the symmetry protected topological phase. Here, $\hat{\ell}_z$ behaves as the symmetry protected topological order. However, the topological phase transition with the spontaneous symmetry breaking takes place at a magnetic field comparable with the dipolar field beyond which $\hat{\ell}_z \!=\! 1$ is realized to minimize the Zeeman magnetic energy. Since the magnetic response becomes isotropic in the high field regime, without loss of generality, the following section will focus on the simple situation where the magnetic field is applied along the surface normal. 

Note that the angle $\varphi$ is locked by minimizing the dipole interaction to~\cite{fujita,Leggett,tewordt,fishman}
\beq
\varphi = \cos^{-1}\left( -\frac{1}{4} \frac{\langle \Delta _{\perp}(z)\Delta _{\parallel}(z)\rangle}{\langle \Delta^2_{\parallel}(z)\rangle} \right), \label{eq:leggett}
\eeq
where we set $\Delta _{\parallel} \!=\! \Delta_x \!=\! \Delta _y$ and $\Delta _{\perp} \!=\! \Delta_z$ and $\langle \cdots\rangle _z \!\equiv\! \frac{1}{D}\int^{D}_0 \cdots dz$. Equation~(\ref{eq:leggett}) depends on the ratio of the pair potentials which are distorted by the nonlinear effect of the Zeeman magnetic field.

\section{Quasiclassical theory for superfluid $^3$He}


The quasiclassical Green's function $\underline{g}$ is obtained from the Nambu-Gor'kov Green's functions $\underline{G}$ with the Matsubara frequency $\omega _n \!=\! (2n+1)\pi T$ ($n\!\in\!\mathbb{Z}$) and the quasiparticle renormalization factor $A$ as
\beq
\underline{g} (\hat{\bm k},{\bm r};i\omega _n) = \frac{1}{A} 
\int^{E _{\rm c}}_{-E _{\rm c}} d\xi _{\bm k}
\underline{\tau}_z \underline{G} ({\bm k},{\bm r}; i\omega _n).
\eeq
The quasiclassical Green's function $\underline{g}$ for spin-triplet superfluids is described in the particle-hole space as
\beq
\underline{g} = \left[
\begin{array}{cc}
g_{0}{\sigma}_{0}+g_{\mu}\sigma _{\mu} & i\sigma _yf_{0}+  i{\sigma}_{\mu} {\sigma}_y f_{\mu} \\
i\sigma _yf^{\dag}_{\rm 0}+ i{\sigma}_y {\sigma}_{\mu} {f}^{\dag}_{\mu} & 
{g}^{\dag}_{0}{\sigma}_{0}+g^{\dag}_{\mu}\sigma^{\ast}_{\mu}
\end{array}
\right],
\eeq
where $f_0\!\equiv\! f_0(\hat{\bm k},{\bm r};i\omega _n)$ and $f_{\mu}\!\equiv\! f_{\mu}(\hat{\bm k},{\bm r};i\omega _n)$ denote the spin-singlet and -triplet components of the quasiclassical Green's function, respectively. Note that $\underline{g}$ satisfies the normalization condition $\underline{g}^2 = -\pi^2 \underline{\tau}_0$. Here, we introduce $\sigma _0$ and $\underline{\tau}_0$ as the unit matrix in spin and particle-hole spaces.

The evolution of the Nambu-Gor'kov Green's functions $\underline{G}$ is governed by the Nambu-Gor'kov equation, $[ -i\omega _n + \underline{\mathcal{H}}]\underline{G} \!=\! \tau _0$. Following the standard procedure, the quasiclassical Green's functions $\underline{g}(\hat{\bm k},{\bm r};i\epsilon _m) $ obeys the so-called Eilenberger equation,\cite{serene}
\beq
&&\hspace{-20mm}
\left[ i\omega _n \underline{\tau}_z-\underline{\mathcal{S}}(\hat{\bm k},{\bm r})-\underline{v}, \underline{g}(\hat{\bm k},{\bm r};i\omega _n) \right] \nn \\
&& 
+ i{\bm v}_{\rm F}(\hat{\bm k})\cdot{\bm \nabla}\underline{g} (\hat{\bm k},{\bm r};i\omega _n) 
= \underline{0}.
\label{eq:eilen}
\eeq
The quasiclassical Green's functions must satisfy a constraint given by the normalization condition, $\underline{g}^2 \!=\! -\pi^2\underline{\tau}_0$. The Fermi velocity ${\bm v}_{\rm F}$ is given as ${\bm v}_{\rm F}(\hat{\bm k}) \!=\! v_{\rm F}\hat{\bm k}$ on the three-dimensional Fermi sphere. The Zeeman magnetic field is included in Eq.~(\ref{eq:eilen}) as 
\beq
\underline{v} \equiv \frac{1}{1+F^{\rm a}_0}\underline{\tau}_z \underline{V} 
= -\frac{1}{1+F^{\rm a}_0} \mu _{\rm n} H_{\mu}\left(\begin{array}{cc} \sigma _{\mu} & 0 \\ 0 & \sigma^{\ast}_{\mu} 
\end{array}\right),
\eeq
where $F^{\rm a}_0$ is one of the Landau parameters which describes the enhancement of the spin susceptibility, as mentioned below. The $4\!\times\! 4$ matrix $\underline{\mathcal{S}}$ describes the quasiclassical self-energies obtained from $\underline{\mathcal{S}}(\hat{\bm k},{\bm r}) \!\approx\! A\underline{\Sigma} ({\bm k}\!=\! k_{\rm F}\hat{\bm k},{\bm r})\underline{\tau}_z$, where $\underline{\Sigma} ({\bm k},{\bm r}) \!=\! \int d{\bm r}_{12}e^{i{\bm k}\cdot{\bm r}_{12}}\underline{\Sigma} ({\bm r}_1,{\bm r}_2)$. The quasiclassical self-energy matrix consists of the Fermi liquid correction in the diagonal elements and the pair potential $d_{\mu}$, 
\beq
\underline{\mathcal{S}}(\hat{\bm k},{\bm r}) = \left[
\begin{array}{cc}
\nu _{0}{\sigma}_{0}+ \nu _{\mu}{\sigma}_{\mu} & 
i{\sigma}_{\mu}{\sigma}_y d_{\mu} \\
i{\sigma}_y{\sigma}_{\mu} d^{\ast}_{\mu} & 
{\nu}^{\prime}_{0}{\sigma}^{\ast}_{0}
+ {\nu}^{\prime}_{\mu}{\sigma}^{\ast}_{\mu}
\end{array}
\right],
\label{eq:s}
\eeq
where we set $\nu _0 \!\equiv\! \nu _{0}(\hat{\bm k},{\bm r})$, $\nu _{\mu}\!\equiv\! \nu _{\mu}(\hat{\bm k},{\bm r})$, and $d_{\mu} \!\equiv\! d_{\mu}(\hat{\bm k},{\bm r})$. We also introduce the notation, $\nu^{\prime}_{0,\mu} \!\equiv\! \nu^{\ast}_{0,\mu}(-\hat{\bm k},{\bm r})$.

The Fermi liquid corrections $\nu _0$ and $\nu _{\mu}$ are associated with the quasiclassical Green's functions ${g}_0$ and $g_{\nu}$ as
\begin{subequations}
\label{eq:nu}
\beq
\nu _{0} (\hat{\bm k},{\bm r}) = \sum _{\ell}A^{({\rm s})}_{\ell}
\left\langle P_{\ell}(\hat{\bm k}\cdot\hat{\bm k}^{\prime}) g_{0}(\hat{\bm k}^{\prime},{\bm r};i\omega _n)
\right\rangle _{\hat{\bm k}^{\prime},n}, 
\eeq
\beq
\nu _{\mu} (\hat{\bm k},{\bm r}) = \sum _{\ell}A^{({\rm a})}_{\ell}
\left\langle P_{\ell}(\hat{\bm k}\cdot\hat{\bm k}^{\prime}) g_{\mu}(\hat{\bm k}^{\prime},{\bm r};i\omega _n)
\right\rangle _{\hat{\bm k}^{\prime},n}, 
\eeq
\end{subequations}
where $\langle \cdots\rangle _{\hat{\bm k},n}$ denotes the Fermi surface average and Matsubara sum: $\langle \cdots\rangle _{\hat{\bm k},n} \!=\! T\sum _{|\omega _n|<E _{\rm c}}\int \frac{d\hat{\bm k}}{4\pi}$. The Fermi liquid corrections are expanded in terms of the Legendre polynomials $P_{\ell}$. The coefficients $A^{({\rm s})}_{\ell}$ and $A^{({\rm a})}_{\ell}$ are the symmetric and antisymmetric quasiparticle scattering amplitudes, which are parametrized with the Landau's Fermi liquid parameters,\cite{serene} $F^{\rm s,a}_{\ell}$, through 
\beq
A^{\rm s,a}_{\ell} = \frac{F^{\rm s,a}_{\ell}}{1+F^{\rm s,a}_{\ell}/(2\ell + 1)}.
\eeq 
The $\ell \!=\! 0$ ($\ell \!=\! 1$) channel of the symmetric part in the Fermi liquid corrections couples to the density distribution (mass current density $J^{\rm m}_{\mu} ({\bm r})$) and the antisymmetric part in the $\ell \!=\! 0$ and $1$ channels arises from the magnetization density $M_{\mu}({\bm r})$ and spin current density $J^{\rm s}_{\mu\nu} ({\bm r})$, respectively. They are defined with the quasiclassical Green's functions $g_0$ and $g_{\mu}$ as 
\begin{subequations}
\beq
M_{\mu} ({\bm r}) = M_{\rm N} \left[ \frac{H_{\mu}}{H} + \frac{1}{\mu _{\rm n}H}
\left\langle g_{\mu}(\hat{\bm k},{\bm r};i\omega _m)
\right\rangle_{\hat{\bm k},\omega_m}  \right], 
\label{eq:M}
\eeq
\beq
J^{\rm m}_{\mu} ({\bm r}) = 2 v_{\rm F} N_{\rm F} 
\left\langle \hat{k}_{\mu}g_0(\hat{\bm k},{\bm r};i\omega _n)\right\rangle_{\hat{\bm k},n},
\eeq
\beq
J^{\rm s}_{\mu\nu} ({\bm r}) = 2 v_{\rm F} N_{\rm F} 
\left\langle \hat{k}_{\nu}g_{\mu}(\hat{\bm k},{\bm r};i\omega _n)\right\rangle _{\hat{\bm k},n}, 
\label{eq:spinflow}
\eeq
\end{subequations}
where the magnetization in the normal state of $^3$He is given by $M_{\rm N} \!=\! 2\mu^2_{\rm n} N_{\rm F} H/(1+F^{\rm a}_0) \!=\! \chi _{\rm N}H$ and $N_{\rm F}$ denotes the density of states at the Fermi energy in a normal Fermi gas. The spin current density $J^{\rm s}_{\mu\nu}$ describes the flow of the spin component $S_{\mu}$ along the $\hat{r}_{\nu}$-direction. As we will emphasize below, the parameter $F^{\rm a}_0$ coupled with the magnetization density strongly affects the qualitative feature of the A-B phase transition induced by a magnetic field. 

The pair potentials $d_{\mu}(\hat{\bm k},{\bm r})$ in spin-triplet superfluids are obtained from the gap equation with an attractive interparticle interaction $V(\hat{\bm k},\hat{\bm k}^{\prime})$, 
\beq
d_{\mu}(\hat{\bm k},{\bm r}) = \left\langle
V(\hat{\bm k},\hat{\bm k}^{\prime}) f_{\mu}(\hat{\bm k},{\bm r}; i\omega _n)
\right\rangle _{\hat{\bm k}^{\prime},n},
\eeq
where the pair interaction $V$ is assumed to be invariant under the $SO(3)_{\bm L} \!\times\! SO(3)_{\bm S}$ rotational symmetry in spin and orbital spaces. Hence, using the form $V(\hat{\bm k},\hat{\bm k}^{\prime}) \!=\! 3g\hat{k}_{\mu}\hat{k}^{\prime}_{\mu}$ with the coupling constant $g \!>\! 0$ and the form of the B-phase order parameter in Eq.~(\ref{eq:op}), the gap equations for $d_{\mu\nu}({\bm r})$ are
\beq
d_{\mu\nu}({\bm r}) = 3g\left\langle
\hat{k}_{\nu} f_{\mu}(\hat{\bm k},{\bm r}; i\omega _n)
\right\rangle _{\hat{\bm k},n}.
\label{eq:gap}
\eeq
The coupling constant $g$ is related to the transition temperature $T_{\rm c0}$ in the bulk, which is given by the linearized gap equation at $T\!=\! T_{\rm c0}$ as
\beq
\frac{1}{g} = \pi T_{\rm c0}\sum _{|\omega _n| \!<\! \omega _{\rm c}} \frac{1}{|\omega _{n,{\rm c}}|},
\eeq 
where $\omega _{\rm c}$ is the cutoff frequency and $\omega _{n,{\rm c}}$ denotes the Matsubara frequency at $T\!=\! T_{\rm c0}$. 

In realistic situation of $^3$He, the magnetic dipole interaction arises from the magnetic moment of $^3$He nuclei which reduces the $SO(3)_{\bm L} \!\times\! SO(3)_{\bm S}$ symmetry to $SO(3)_{{\bm L}+{\bm S}}$. In the presence of a perpendicular magnetic field which we consider here, however, as discussed in Sec.~II, the dipole interaction merely locks the angle $\varphi$ to Eq.~(\ref{eq:leggett}) and the contribution to the thermodynamics is negligible. Hence, for a perpendicular field, the order parameter reduces to 
\beq
d_{\mu\nu}({\bm r}) = \delta _{\mu\nu} \Delta _{\nu}({\bm r}),
\eeq 
which corresponds to the case of $\hat{\bm n} \!\parallel\! \hat{\bm z}$ in Eq.~(\ref{eq:op}). The effect of the dipole interaction becomes crucial in the case of a weak magnetic field parallel to the surface, which will be discussed elsewhere.~\cite{TM2012,TM2012dd}

In summary, the Eilenberger equation (\ref{eq:eilen}) coupled with Eqs.~(\ref{eq:nu}) and (\ref{eq:gap}) through the quasiclassical self-energies in Eq.~(\ref{eq:s}) provides the closed form of the self-consistent equations for the quasiclassical Green's functions $\underline{g}$ and the meanfield potentials $\nu_0$, $\nu _{\mu}$, and $d_{\mu}$. In Appendix B, we describe in details the calculated systems, boundary conditions, and the procedure for numerical calculations. 

\begin{figure}[b!]
\includegraphics[width=50mm]{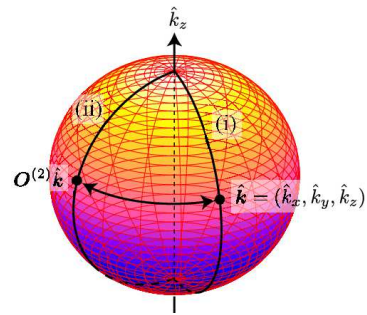}
\caption{(Color online) One-to-one correspondence between two points on the three-dimensional Fermi sphere. The quasiclassical Green's function at an arbitrary $O\hat{\bm k}$ belonging to the path (ii) is obtained by the $SO(2)_{L_z+S_z}$ rotation of $\underline{g}(\hat{\bm k},z;i\omega _n)$ calculated on the path (i).}
\label{fig:sphere}
\end{figure}

As shown in Appendix C, the Eilenberger equation (\ref{eq:eilen}) for $^3$He-B is invariant under the $SO(2)_{L_z+S_z}$ rotation, when the Zeeman magnetic field is applied along the surface normal. The symmetry leads to the one-to-one correspondence of the quasiclassical Green's function between two points $\hat{\bm k}$ and $O^{(2)}\hat{\bm k}$ on the Fermi sphere,
\beq
\underline{g}(O^{(2)}\hat{\bm k},z; i\omega _n) = \underline{\mathcal{U}}^{\dag}_2 \underline{g}(\hat{\bm k},z; i\omega _n)\underline{\mathcal{U}}_2. 
\label{eq:Gso2main}
\eeq
where $O^{(2)}$ is an $SO(2)$ rotation matrix about the ${\bm z}$-axis and $\mathcal{U}_2$ is the $4\!\times\!4$ matrix which describes an $SU(2)$ rotation associated with $O^{(2)}$ (for the details, see Appendix C). This relation through the $SO(2)$ rotation is useful for shorting the computation time of the selfconsistent calculation. Once we calculate $\underline{g}(\hat{\bm k},z; i\omega _n)$ along the path (i) displayed in Fig.~\ref{fig:sphere}, the Green's function $\underline{g}$ for all $\hat{\bm k}$ is given by the symmetric relation in (\ref{eq:Gso2main}) with $\underline{g}(\hat{\bm k},z; i\omega _n)$. 

Throughout this paper, we use the set of the Fermi liquid parameters, $F^{\rm s}_{0}\!=\! 9.3$, $F^{\rm a}_0 \!=\! -0.695$, $F^{\rm s}_1 \!=\! 5.39$, and $F^{\rm a}_1 \!=\! -0.5$.~\cite{vollhardt} The cutoff frequency on the Matsubara sum is taken to be $\omega _{\rm c} \!=\! 20\pi T_{\rm c0}$ for low temperatures and $160\pi T_{\rm c0}$ for high temperatures. All length and energy scales are in a unit of the coherence length in quasiclassical formalism, $\xi _0 \!=\! v_{\rm F}/\pi T_{\rm c0}$, and $\pi T_{\rm c0}$.

\section{Order parameters, local spin susceptibilities, and surface bound states}

\subsection{Distortion of the B-phase order parameter and magnetization}

\begin{figure}[bt!]
\includegraphics[width=70mm]{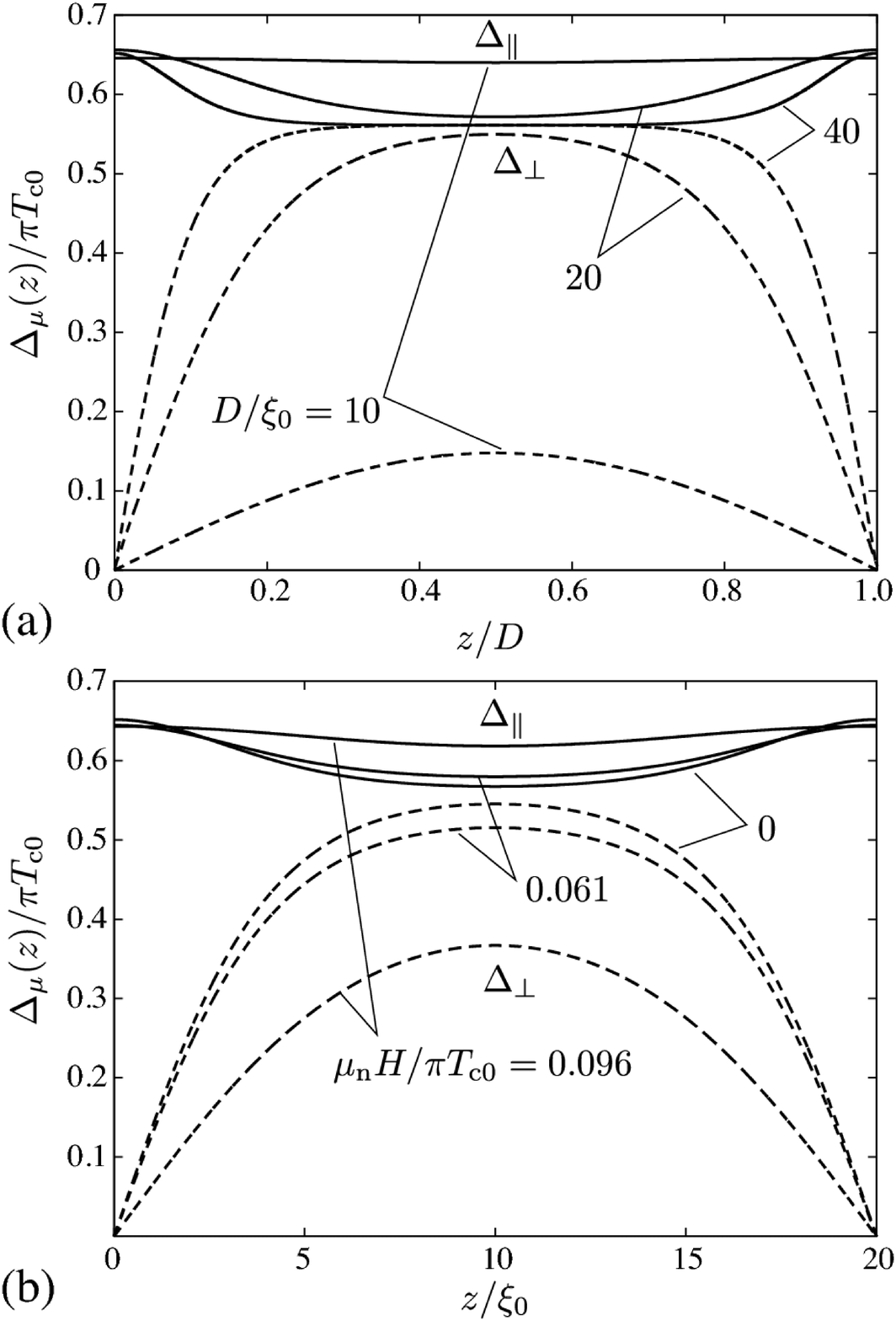}
\caption{(a) Spatial profiles of $\Delta _{\parallel}(z)$ (solid lines) and $\Delta _{\perp}(z)$ (dashed lines) for $D/\xi _0 \!=\! 10$, $20$, and $40$ at $H \!=\! 0$ and $T\!=\!0.2T_{\rm c0}$. The horizontal axis is scaled with $D$. (b) Field dependence of $\Delta _{\parallel}(z)$ (solid lines) and $\Delta _{\perp}(z)$ (dashed lines) for $D  \!=\! 20 \xi _0$ and $T\!=\! 0.2T_{\rm c0}$, where $\mu _{\rm n}H /\pi T_{\rm c0}\!=\! 0$, $0.061$, and $0.096$.}
\label{fig:gap}
\end{figure}

First of all, in Fig.~\ref{fig:gap}(a), we summarize the spatial profiles of $\Delta _{\parallel}(z)$ and $\Delta _{\perp}(z)$ for various thickness $D$ where we fix $H \!=\! 0$ and $T\!=\!0.2T_{\rm c0}$. In the vicinity of the specular surface, the $\Delta _z$ component is suppressed by the pair breaking effect and the parallel components remain isotropic, that is, $\Delta _x \!=\! \Delta _y \!\equiv \! \Delta _{\parallel}$ and $\Delta _z \!\equiv\! \Delta _{\perp}$. For a large $D$, {\it e.g.}, $D\!=\! 40\xi _0$, the isotropic B-phase order parameter with $\Delta _{\parallel} \!=\! \Delta _{\perp}$ appears around the middle region $z/D \!\sim\! 0.5$. It is continuously turned to the planar state with $\Delta _z \!=\! 0$ at $z \!=\! 0$ across the squashed B-phase with $\Delta _{\parallel} \!>\! \Delta _z$. As $D$ decreases, the pair breaking effect at the surface occurs even in the central region, which elliptically squeezes the order parameters. The squashed B-phase undergoes a second-order phase transition to the planar or A-phase at the thickness $D \!\approx\! 9.6 \xi _0 \!\equiv\! D_{\rm cri}(H\!=\! 0)$,~\cite{haraJLTP1988,vorontsov} when the magnetic field is absent.

Figure~\ref{fig:gap}(b) shows the spatial profiles of $\Delta _{\parallel}(z)$ and $\Delta _{\perp}(z)$ for $\mu _{\rm n}H /\pi T_{\rm c0}\!=\! 0$, $0.061$, and $0.096$ and $D\!=\! 20\xi _0$. It is seen that the magnetic field ${\bm H} \!\parallel\! \hat{\bm z}$ as well as thickness $D$ squeezes the B-phase order parameter elliptically, leading to $\Delta _{\parallel} \!>\! \Delta _{\perp}$. 
As we will mention in the subsequent section, the squashed B-phase undergoes the first-order phase transition to the planar or A-phase at low temperature. 

\begin{figure}[tb!]
\includegraphics[width=70mm]{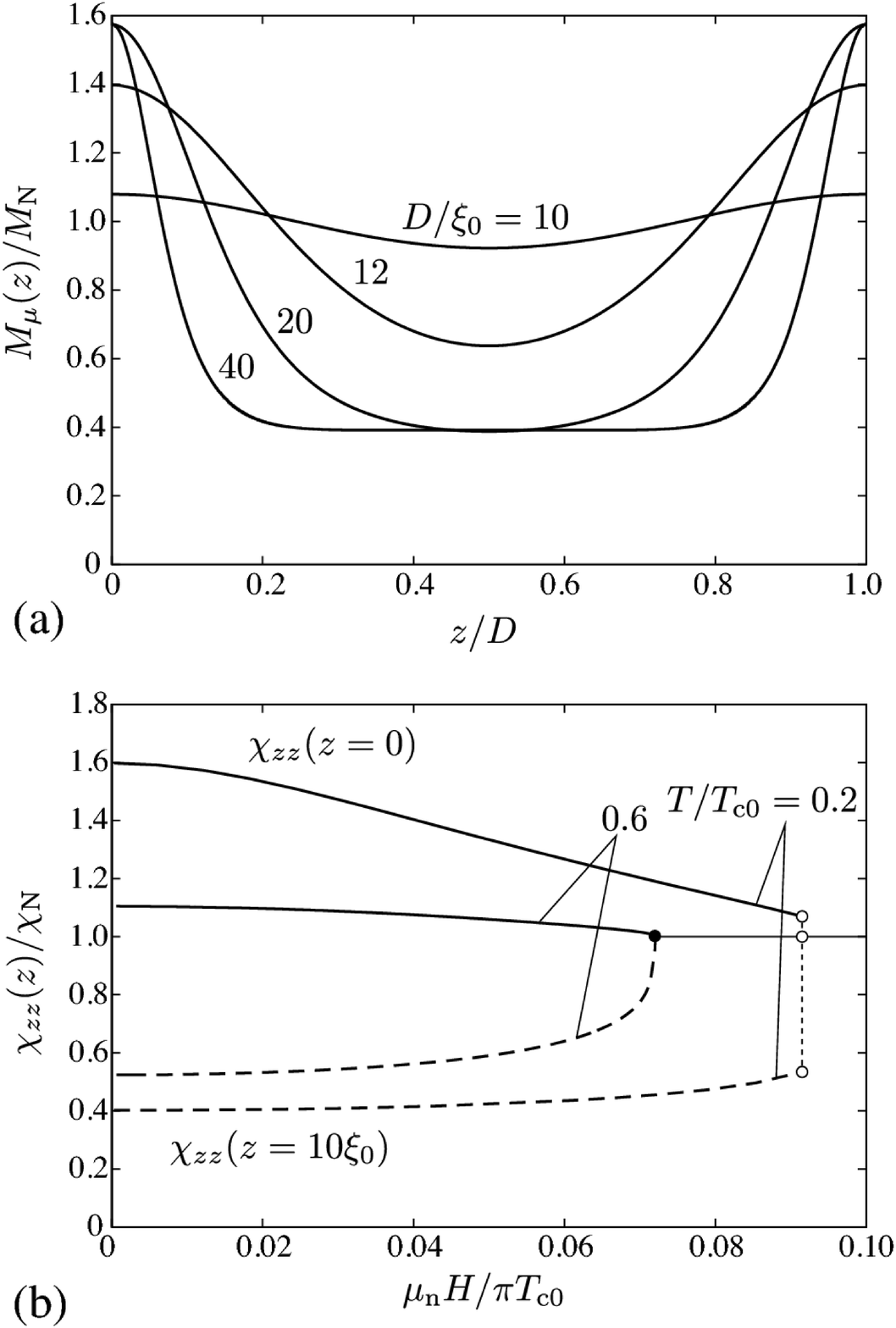}
\caption{(a) Spatial profiles of the local magnetization density $M_{\mu}(z)/M_{\rm N}$, corresponding to the ratio of the local spin susceptibility $\chi _{zz}(z)/\chi _{\rm N}$, for various $D$'s at $T\!=\! 0.2T_{\rm c0}$ and $\mu _{\rm n}H/\pi T_{\rm c0} \!=\! 0.0122$. (b) Field-dependence of $\chi _{zz}(z)$ on the surface $z \!=\! 0$ (solid line) and $z \!=\! 10\xi _0$ (dashed line) at $T \!=\! 0.2 T_{\rm c0}$ and $0.6 T_{\rm c0}$ where $D \!=\! 20 \xi _0$ is fixed.}
\label{fig:mag}
\end{figure}

The thickness-dependence of the local magnetization density $M_{\mu}(z)/M_{\rm N}$ defined in Eq.~(\ref{eq:M}) is summarized in Fig.~\ref{fig:mag}(a). In the case of a large $D/\xi _0$, the magnetization around the central region, {\it e.g.}, $z/D \!=\! 0.5$, is strongly suppressed, compared with that in the normal $^3$He. In the thermodynamic limit, $D\!\rightarrow\! \infty$, the ratio of the magnetization between the B-phase and normal phase is obtained as $\chi _{zz}/\chi _{\rm N} \!=\! 2(1+F^{\rm a}_0)/(3+2F^{\rm a}_0)$ at $T\!=\! 0$,~\cite{vollhardt,schopohl}, which implies that $\chi _{zz} \!\approx\! 0.38\chi _{\rm N}$ for $F^{\rm a}_0 \!=\! -0.695$. It is important to mention that for a large $D/\xi _0$, a magnetic field perpendicular to the surface enhances the low-temperature spin susceptibility on the surface, where $M_{z}(z\!=\! 0) \!>\! M_{\rm N}$ and $M_{z}(z \!\sim\! D/2) \!<\! M_{\rm N}$. The enhancement is closely associated with the energy spectrum of the surface bound states, which will be clarified in the subsequent subsection. As $D$ approaches the critical value $D_{\rm cri}(0)$, the B-phase continuously changes to the planar phase through the squashed B-phase, where the spin susceptibility in the planar phase is indistinguishable from that in the normal state, $\chi^{\rm planar}_z\!=\! \chi _{\rm N}$. Hence, in this regime, the enhancement ceases to exist and the magnetization density flattens due to the strong distortion of the B-phase order parameter. 

The local magnetization density feedbacks the effective magnetic field through the Fermi liquid corrections. This gives rise to a nonlinear effect of the Zeeman magnetic field. Since the distorted B-phase is not accompanied by the mass flow, the quasiclassical selfenergies $\nu(\hat{\bm k},{\bm r})$ are composed of the local magnetization density $M_{\mu}({\bm r})$ and the superfluid spin flow $J^{\rm s}_{\mu\nu}({\bm r})$, which changes the Zeeman energy term to 
\beq
\left[-\frac{\mu _{\rm n}H_{\mu}}{1+F^{\rm a}_0} + \nu _{\mu} (\hat{\bm k},{\bm r}) \right]\sigma _{\mu}
\equiv -\frac{\mu _{\rm n}}{1+F^{\rm a}_0} H^{\rm eff}_{\mu}(\hat{\bm k},{\bm r})\sigma _{\mu},
\eeq
where $H^{\rm eff}_{\mu}(\hat{\bm k},{\bm r})$ denotes the magnetic field deviated by the Fermi liquid corrections, 
\beq
H^{\rm eff}_{\mu}(\hat{\bm k},{\bm r}) &=& H \bigg[ \hat{h}_{\mu}
+ F^{\rm a}_0\left\{ \hat{h}_{\mu} - \frac{M_{\mu}({\bm r})}{M_{\rm N}}\right\}\bigg] \nn \\
&& - \frac{3(1+F^{\rm a}_0)F^{\rm a}_1}{2(3+F^{\rm a}_1)\mu _{\rm n}  v_{\rm F}N_{\rm F}} J^{\rm s}_{\mu\nu}({\bm r}) \hat{k}_{\nu} .
\label{eq:Heff}
\eeq
For realistic situation with $F^{\rm a}_0 \!<\! 0$, the enhancement of the surface magnetization $M_{z}(0)\!>\! M_{\rm N}$ increases the effective magnetic field, $H^{\rm eff}_z (\hat{\bm k},{\bm r}) \!>\! H$, while the suppression of the magnetization in the middle region leads to $H^{\rm eff}_z(\hat{\bm k},{\bm r}) \!<\! H$.

The field-dependence of the local spin susceptibility $M_{z}(z)/M_{\rm N}$ at finite temperatures is summarized in Fig.~\ref{fig:mag}(b). This ratio corresponds to the local spin susceptibilities $\chi _{zz}(z)$ for an applied field ${\bm H}\!\parallel\! \hat{r}_{\nu}$, 
\beq
\frac{\chi _{\mu \nu}(z)}{\chi _{\rm N}} \equiv \frac{M_{\mu}(z)}{M_{\rm N}}.
\label{eq:local_chi}
\eeq
The spin susceptibility in the normal $^3$He is obtained from $\chi _{\rm N} \!\equiv\!  M _{\rm N}/H \!=\! 2\mu^2_{\rm n}N_{\rm F}/(1+F^{\rm a}_0)$. In the low temperature regime, such as $T\!=\! 0.2T_{\rm c0}$, the spin susceptibilities at the surface, $\chi _{zz}(0)/\chi _{\rm N}$, is enhanced in the linear regime of the magnetic field. As  $H$ increases, however, it reduces to $\chi _{\rm N}$ at the surface, while $\chi _{zz}(D/2)$ is insensitive to $H$ as a result of the first-order transition from the B- to A- (or planar) phase. Although the nonlinear effect of the magnetic field suppresses the enhancement of the spin susceptibility due to the SABS at low temperatures, as seen in Fig.~\ref{fig:mag}(b), the spin susceptibility at the surface is still distinct from that in the central region of the system. In the higher temperature region where the B-phase undergoes the second-order transition to the planar or A-phase, $\chi _{zz}(0)$ decreases to $\chi _{N}$ and $\chi _{zz}(D/2)$ gradually increases as $H$ increases.

\subsection{Relation between surface bound states and enhancement of magnetization}

As shown in Eq.~(\ref{eq:sabs}), a Zeeman magnetic field perpendicular to the surface opens a finite energy gap. As shown in Fig~\ref{fig:mag}, it simultaneously induces a large amount of the magnetization at the surface. Here, we clarify the relation between gapped surface bound states and the enhancement of the magnetization density. 

\begin{figure}[bt!]
\includegraphics[width=80mm]{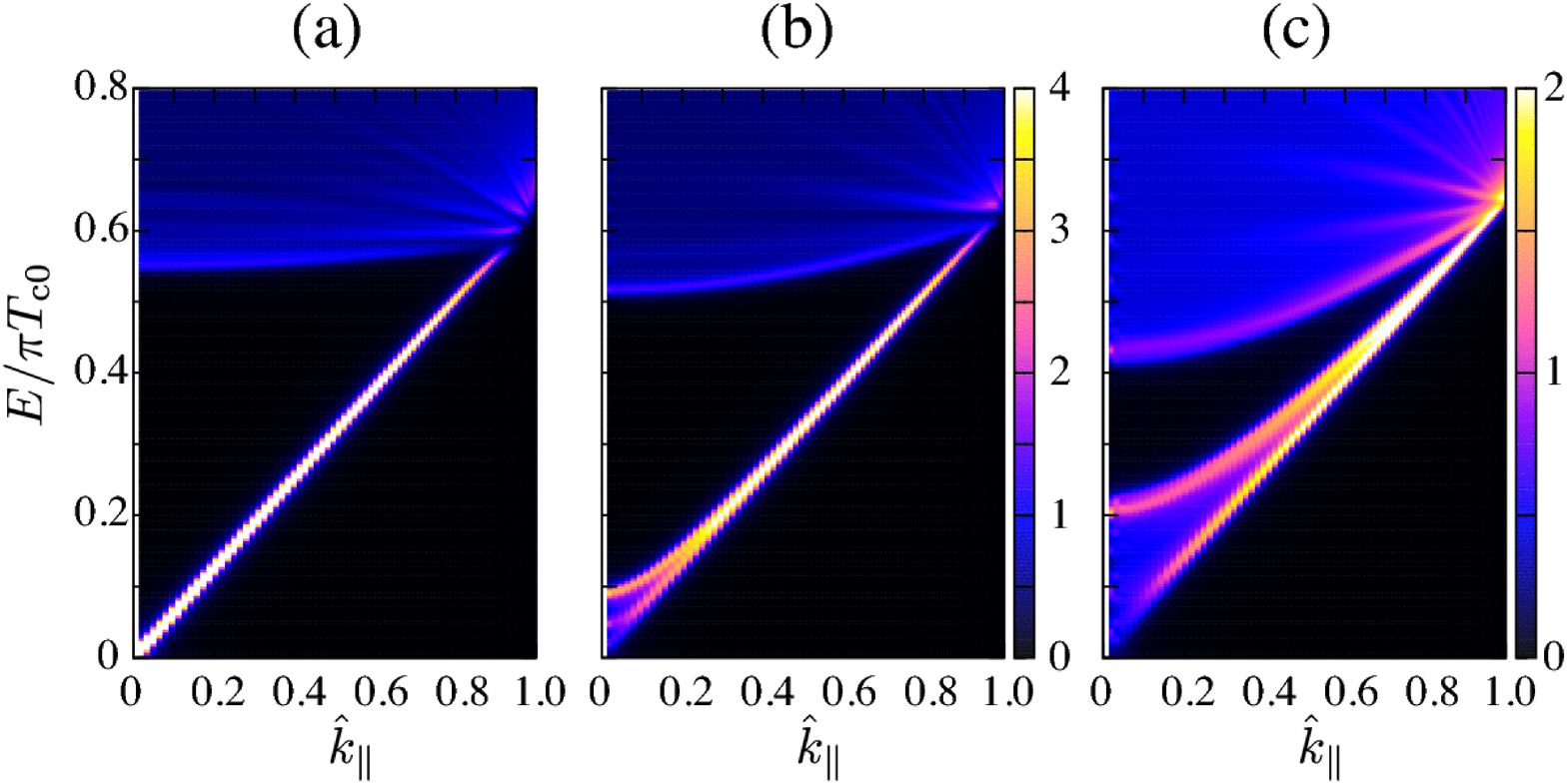}
\includegraphics[width=80mm]{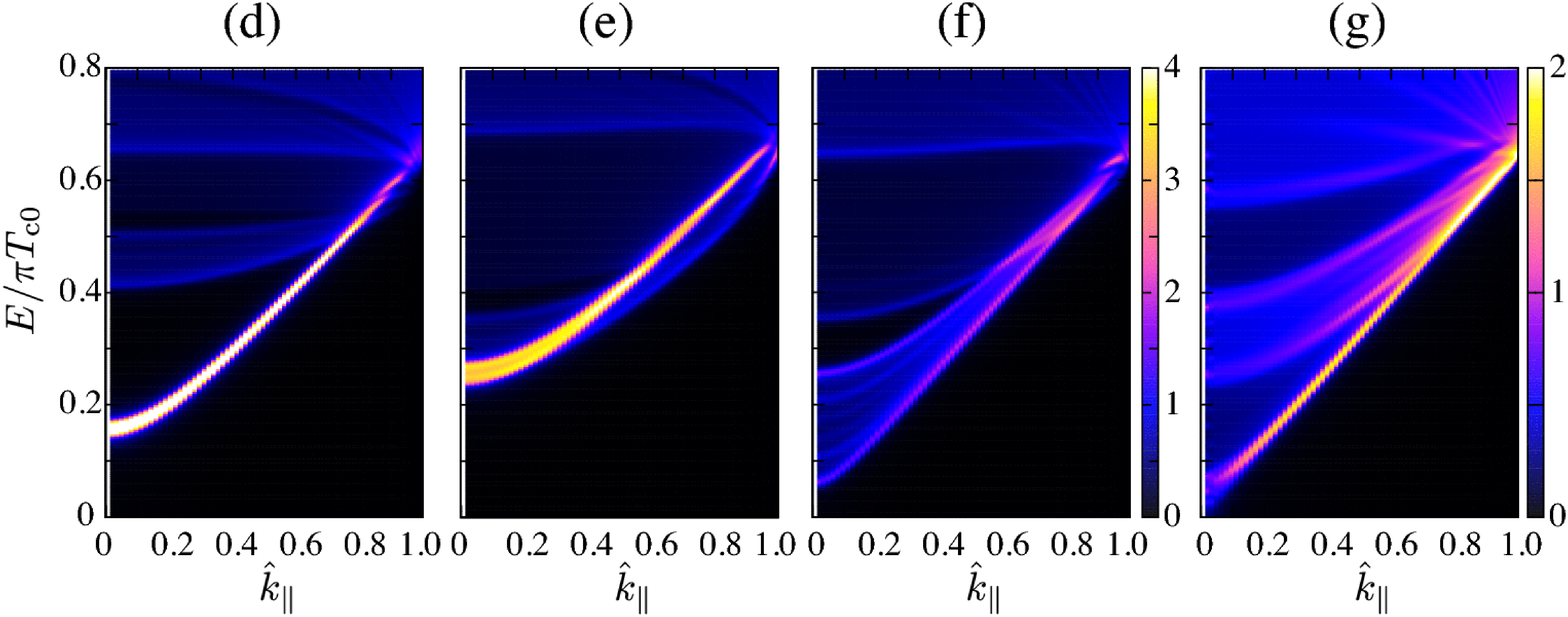}
\caption{(Color online) $\hat{\bm k}$-resolved surface density of states $\mathcal{N}(\hat{k}_{\parallel}, z \!=\! 0,E)$ for $D \!=\! 20 \xi _0$ (a), $12\xi _0$ (b), and $10\xi _0$ (c), where $H\!=\! 0$. $\mathcal{N}(\hat{k}_{\parallel}, z \!=\! 0,E)$ for $D\!=\! 20\xi_0$ at $\mu _{\rm n}H \!=\! 0.0488\pi T_{\rm c0}$ (d) and $0.0854\pi T_{\rm c0}$ (e). (f) and (g) are for $D\!=\! 12\xi _0$ and $10 \xi _0$ at $\mu _{\rm n}H \!=\! 0.0488\pi T_{\rm c0}$. In all the data, the temperature is set to be $T \!=\! 0.2T_{\rm c0}$.}
\label{fig:sdos}
\end{figure}

First, in Figs.~\ref{fig:sdos}(a)-(c), we display the $\hat{\bm k}$-resolved surface density of states, 
\beq
\mathcal{N}(\hat{k}_{\parallel}, {\bm r},E) = - \frac{1}{\pi} {\rm Im} \int^{2\pi}_0\frac{d\phi _{\bm k}}{2\pi}
g^{\rm R}_0 (\hat{\bm k},{\bm r};E) ,
\label{eq:ldosk} 
\eeq
when the magnetic field is absent. In Eq.~(\ref{eq:ldosk}), $\hat{k}_{\parallel}$ denotes the momentum parallel to the surface, $\hat{k}_{\parallel} \!=\! \sin \theta _{\bm k}$, as shown in Fig.~\ref{fig:system}. The retarded Green's function $g^{\rm R}_0 (\hat{\bm k},{\bm r};E)$ is obtained from Eq.~(\ref{eq:eilen}) with $i\omega _n \rightarrow E+i0_+$. Throughout this paper, we fix $0_+ \!\equiv\! 0.005 \pi T_{\rm c0}$. Since the squashed B-phase in a slab geometry is $SO(2)_{{\bm L}+{\bm S}}$ symmetric around the $\hat{\bm z}$-axis, $\mathcal{N}(\hat{k}_{\parallel}, {\bm r},E)$ describes the dispersion relation of the surface bound state.

In the absence of a Zeeman magnetic field, the time-reversal symmetry as well as the particle-hole is preserved. Hence, the BdG Hamiltonian $\underline{\mathcal{H}}({\bm k})$ is anticommutable with the chiral operator $\underline{\Gamma}$ combined with the time-reversal operator $\underline{\mathcal{T}} \!=\! i\sigma _y \underline{\tau}_0 K$ and particle-hole operations $\underline{\mathcal{C}} \!=\! \sigma _x \underline{\tau}_yK$, which is called the chiral symmetry, $\{ \underline{\mathcal{H}}({\bm k}),\underline{\Gamma} \} \!=\! 0$. Here, $K$ is the complex conjugate operator. The chiral symmetry allows one to introduce a three-dimensional winding number, $w \!=\! \int \frac{d{\bm k}}{24\pi^2}\epsilon _{\mu\nu\eta} {\rm Tr}[\underline{\Gamma}(\underline{\mathcal{H}}^{-1}\partial _{\mu}\underline{\mathcal{H}})(\underline{\mathcal{H}}^{-1}\partial _{\nu}\underline{\mathcal{H}})(\underline{\mathcal{H}}^{-1}\partial _{\eta}\underline{\mathcal{H}})] $, which is evaluated as $w \!=\! 2$ for the B-phase.~\cite{schnyder,volovik2009} Hence, the B-phase in the absence of a magnetic field is a topological phase and the bulk-edge correspondence implies the SABS satisfies $E({\bm k}_{\parallel}) \!=\! 0$ at ${\bm k}_{\parallel} \!=\! {\bm 0}$, which is consistent with the analytic solution of the BdG equation within the Andreev approximation. 

The $\hat{\bm k}$-resolved surface density of states for $D \!=\! 20\xi_0$ without a magnetic field, which is displayed in Fig.~\ref{fig:sdos}(a), is consistent with the topological consideration, where the gapless point exists at ${\bm k}_{\parallel} \!=\! {\bm 0}$. However, since the SABS is localized at the surface within the coherence length scale $\xi _0$, the wavefunctions at both two surfaces are overlapped with each other as the thickness $D$ approaches $\xi _0$. As discussed in Ref.~\onlinecite{tsutsumiPRB,TM2010,TM2010v2,kawakami,cheng1,cheng2}, the hybridization of wavefunctions localized at $z\!=\! 0$ and $D$ split the gapless cone as $e^{-D/\xi}$. Indeed, as seen in Figs.~\ref{fig:sdos}(b) and \ref{fig:sdos}(c), the spectral weight at ${\bm k}_{\parallel} \!=\! {\bm 0}$ weakens as the thickness $D$ approaches $D_{\rm cri}(0) \!=\! 9.6\xi _0$. In addition, it has the double peak in the low energy region, where the upper branch has a distinct energy gap at ${\bm k} _{\parallel} \!=\! {\bm 0}$ and another one remains almost linear at finite $\hat{k}_{\parallel}$. For $D \!=\! 10\xi _0$, the upper branch which has a energy gap $E\!=\! 0.2\pi T_{\rm c0}$ originates from the hybridization of Majorana cones bound at two surfaces, while the lower branch reflects the fact that the pair potential $\Delta _{\perp}$ which is perpendicular to the surface is squashed by two specular surfaces as displayed in Fig.~\ref{fig:gap}. At $D \!=\! D_{\rm cri}(0)$, the squashed B-phase order parameter continuously turns to the planar phase with $\Delta _{\perp} \!=\! 0$ where $\hat{\bm k}_{\parallel} \!=\! {\bm 0}$ corresponds to the location of the point nodes in the bulk. The planar phase, the point node of which is normal to the surface, is not accompanied by the surface bound state and the low energy spectrum is linear on $\hat{k}_{\parallel}$ in the whole system. 


As seen in Fig.~\ref{fig:sdos}(d), the perpendicular field opens a finite energy gap in the surface cone, $\min|E| \!\sim\! 0.15\pi T_{\rm c0}$. For $\mu _{\rm n}H \!=\! 0.0488\pi T_{\rm c0}$ and $T \!=\! 0.2 T_{\rm c0}$, it is seen in Fig.~\ref{fig:mag} that the value of $M_{z}(z)/M_{\rm N}$ at the surface $z \!=\! 0$ is about $1.4$. Then, the effective Zeeman energy at the surface $z \!=\! 0$ is estimated from Eq.~(\ref{eq:Heff}) as $\mu _{\rm n} H^{\rm eff}_{z}(\hat{\bm k},{\bm r})/(1+F^{\rm a}_0) \!\approx \! 0.2\pi T_{\rm c0}$. At $z \!=\! 10 \xi _0$, however, it decreases to $0.1 \pi T_{\rm c0}$, because of the suppression of the spin susceptibility $M_z(z \!=\! 10 \xi_0)/M_{\rm N} \!\approx\! 0.4$. Hence, the energy gap $\min|E| \!\sim\! 0.15\pi T_{\rm c0}$ in Fig.~\ref{fig:sdos}(d) is approximately consistent with the analytic dispersion in Eq.~(\ref{eq:sabs}) with the spatially averaged effective Zeeman energy. In the high magnetic field (Fig.~\ref{fig:sdos}(e)), however, the nonlinear effect of the Zeeman magnetic field causes the pair breaking effect as displayed in Fig.~\ref{fig:gap}(b). Therefore, as $H$ increases, the bulk excitation gap becomes lower in addition to the increase of the energy gap of surface bound state. This behavior is confirmed in Fig.~\ref{fig:sdos}(e) where the continuous excitation band lowers and merges to the gapped SABS branch. Since the situations of $D \!=\! 12\xi _0$ and $10\xi _0$ in Figs.~\ref{fig:sdos}(f) and \ref{fig:sdos}(g) are close to the second-order phase transition field, the surface cone ceases to exist and the quasiparticle excitations in the entire system become gapless. 

\begin{figure}[tb!]
\includegraphics[width=70mm]{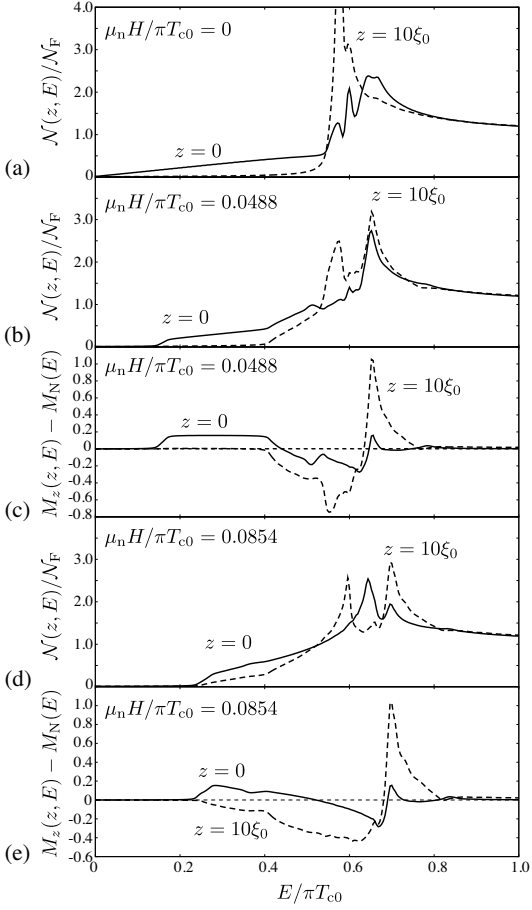}
\caption{Local density of states $\mathcal{N}(z,E)$ at $z \!=\! 0$ (the solid line) and $z \!=\! 10 \xi _0$ (the dashed line) for $\mu _{\rm n} H \!=\! 0$ (a), $0.0488\pi T_{\rm c0}$ (b), and $0.0854\pi T_{\rm c0}$ (d), where ${\bm H}$ is perpendicular to the surface. (c,e) $-{\rm Im}\langle g^{\rm R}_{\mu}(\hat{\bm k},z; E) \rangle _{\hat{\bm k}}$ in the same condition as (b,d). In all the data, the temperature and thickness are set to be $T \!=\! 0.2T_{\rm c0}$ and $D\!=\! 20\xi_0$.}
\label{fig:ldosz}
\end{figure}

Now, let us clarify how the change of the spectrum of the surface bound states affects the local magnetization density at the surface. In Figs.~\ref{fig:ldosz}(a), \ref{fig:ldosz}(b), \ref{fig:ldosz}(d), we plot the surface density of states $\mathcal{N}(z \!=\! 0,E)$ in the absence and presence of a perpendicular magnetic field, respectively, which correspond to Fig.~\ref{fig:sdos}(a), \ref{fig:sdos}(d), and ~\ref{fig:sdos}(e). The local density of states $\mathcal{N}({\bm r},E)$ is defined as 
\beq
\mathcal{N}({\bm r},E) &=& - \frac{1}{\pi} {\rm Im} \left\langle g^{\rm R}_0 (\hat{\bm k},{\bm r};E) \right\rangle _{\hat{\bm k}} \nn \\
&=& \frac{1}{2}\int^{\pi}_0 d\theta _{\bm k}\sin\theta _{\bm k} \mathcal{N}(\hat{k}_{\parallel}, {\bm r}, E),
\eeq
where $\langle \cdots \rangle _{\hat{\bm k}}$ denotes the average on the three-dimensional Fermi surface. The surface density of states in the absence of a magnetic field displayed in Fig.~\ref{fig:ldosz}(a) is linear on $E$ in the low energy region. The linear dependence reflects the gapless cone $E ({\bm k}) \!\propto\! \sqrt{k^2_x+k^2_y}$ and is distinguishable from the full gap behavior at $z \!=\! 10\xi_0$. In the case of ${\bm H}\!\parallel\! \hat{\bm z}$, $\mathcal{N}(z\!=\! 0,E)$ is accompanied by the finite energy gap within $|E| \!\lesssim\! 0.15\pi T_{\rm c0}$. In the presence of a perpendicular Zeeman field, as seen in Figs.~\ref{fig:ldosz}(b) and \ref{fig:ldosz}(d), the finite energy gap appears in the low $E$ region. In the high field regime, the surface density of states loses the linearity in the low $E$ region and is indistinguishable from $\mathcal{N}(z \!=\! D/2,E)$, due to the distortion of the order parameter induced by the nonlinear Zeeman effect.

\begin{figure}[tb!]
\includegraphics[width=70mm]{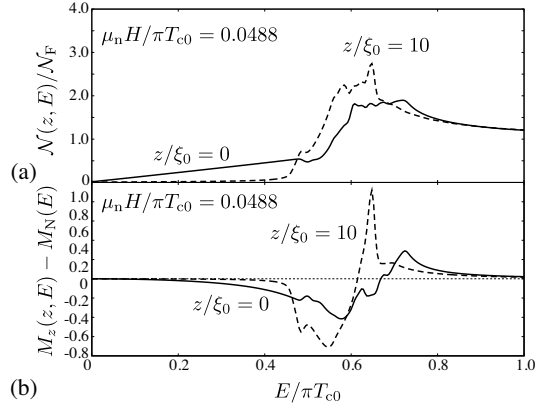}
\caption{Local density of states $\mathcal{N}(z,E)$ at $z \!=\! 0$ (the solid line) and $z \!=\! 10 \xi _0$ (the dashed line) for $\mu _{\rm n} H \!=\! 0.0488\pi T_{\rm c0}$ (b), where ${\bm H}\!\parallel\! \hat{\bm x}$ and $\hat{\bm n}\!\parallel\! \hat{\bm z}$. (b) $-{\rm Im}\langle g^{\rm R}_{\mu}(\hat{\bm k},z; E) \rangle _{\hat{\bm k}}$ for the same situation as (a). The other parameters are same as those in Fig.~\ref{fig:ldosz}.}
\label{fig:ldosx}
\end{figure}

For comparison, we present in Fig.~\ref{fig:ldosx}(a) the surface density of states with the gapless dispersion of the SABS under a magnetic field. This is realized when the magnetic field is parallel to the surface (${\bm H}\!\parallel\! \hat{\bm x}$) and $\hat{\bm n}$ is fixed to be normal to the surface, $\hat{\bm n}\!\parallel\!\hat{\bm z}$, which corresponds to $\hat{\ell}_z  \!=\! R_{xz}(\hat{\bm n} \!=\! \hat{\bm z},\varphi) \!=\! 0$ in Eq.~(\ref{eq:sabs}). Hence, the surface density of states for ${\bm H} \!\parallel\! \hat{\bm x}$ remains linear on $E$. Note that the configuration of $\hat{\bm n}\!\parallel\!\hat{\bm z}$ becomes energetically unstable in the strong magnetic field regime,~\cite{TM2012} because the ground state has the $\hat{\bm n}$-vector texture which satisfies the condition $\hat{\ell}_z (\hat{\bm n},\varphi) \!\equiv\! R_{xz}(\hat{\bm n},\varphi) \!=\! 1$. The ground state under a strong parallel field is necessarily accompanied by the Zeeman energy gap of the surface bound state.

Then, we introduce ${\rm Im}\langle g^{\rm R}_{\mu}(\hat{\bm k},z; E) \rangle _{\hat{\bm k}}$, which is associated with the contribution of quasiparticles in the superfluid state to the local magnetization density $M_{\mu}(z)$, that is,
\beq
M_{\mu}(z) - M_{\rm N}\hat{h}_{\mu} = -\frac{M_{\rm N}}{\mu _{\rm n}H} \int dE {\rm Im}\left\langle g^{\rm R}_{\mu}(\hat{\bm k},z; E) \right\rangle _{\hat{\bm k}}.
\eeq
As seen from Fig.~\ref{fig:ldosz}(c) with the solid line, the quantity $-{\rm Im}\langle g^{\rm R}_{\mu}(\hat{\bm k},z; E) \rangle _{\hat{\bm k}}$ at the surface $z \!=\! 0$ becomes positive in the energy region lower than the bulk excitation gap, $E \!\lesssim\! 0.6\pi T_{\rm c0}$. This implies that the gapped surface bound state considerably enhances the local magnetization density, compared with $M_{\rm N}$. As the Zeeman magnetic field increases, however, the gapped surface bound state merges to the bulk excitations as seen in Fig.~\ref{fig:sdos}(e) and the positive contribution to $M_{z}$ decreases. This is seen in Fig.~\ref{fig:ldosz}(e) with the solid line. In the nonlinear regime of $H$, the resulting magnetization density at the surface becomes comparable with that in the normal $^3$He. Note that since even within this regime the magnetization in the central region of the system ($z \!=\! D/2$) stays around $M_z \!\sim\! 0.4 M_{\rm N}$, the enhancement of the magnetization at the surface is still distinguishable from that in the bulk. 

This is in contrast to the case of the parallel magnetic field where the surface Majorana cone is assumed to remain gapless. It is demonstrated in Fig.~\ref{fig:ldosx}(b) that the low energy quasiparticles in the gapless Majorana cone, $|E|\!\lesssim\!0.4\pi T_{\rm c0}$, do not contribute to the magnetization. The contribution of $-{\rm Im}\langle g^{\rm R}_{\mu}(\hat{\bm k},z; E) \rangle _{\hat{\bm k}}$ in the higher energy region $E \!\gtrsim\! 0.6\pi T_{\rm c0}$ comes up to the negative value, which suppresses the magnetization density relative to $M_{\rm N}$. Hence, as long as the $\hat{\bm n}$-vector is polarized to the surface normal, the magnetization density at the surface becomes highly anisotropic, which is associated with the dispersion of the surface bound state. This consequence is consistent with the interpretation of the surface bound states as the Majorana Ising spin.

\section{Phase diagram and spatially averaged spin susceptibility}

In this section, we present the superfluid phase diagram of $^3$He in a restricted geometry and the $H$- and $T$-dependencies of spin susceptibility averaged over the slab, where the latter is associated with the NMR frequency shift and absorption. In order to discuss the thermodynamic stability and the phase diagram of superfluid $^3$He in a slab geometry, we estimate the thermodynamic functional within the quasiclassical approximation, 
\beq
\delta \Omega [\underline{g}]
= \frac{1}{2} \int^{1}_0 d\lambda {\rm Sp}^{\prime} \left\{ \underline{\nu}
\left( \underline{g}_{\lambda} - \frac{1}{2}\underline{g}  \right)
\right\} ,
\label{eq:omega}
\eeq
where we set 
\beq
{\rm Sp}^{\prime}\{ \cdots\} 
= N_{\rm F} \int d{\bm r} \langle {\rm Tr}_4 \{ \cdots\} \rangle _{\hat{\bm k},\omega _n}. 
\eeq
The quasiclassical auxiliary function $\underline{g}_{\lambda}$ is obtained from the quasiclassical Eilenberger equation (\ref{eq:eilen}) with replacing $\underline{\nu}\!\rightarrow\! \lambda \underline{\nu}$ ($\lambda \!\in\![0,1]$), where the equation is solved once under a given self-energy but not self-consistently. The functional in Eq.~(\ref{eq:omega}) is obtained from the Luttinger-Ward thermodynamic functional associated with the Nambu-Gor'kov Green's function $\underline{G}$, the detailed derivation of which is followed by the work in Ref.~\onlinecite{vorontsov}. Equation (\ref{eq:omega}) includes the influence of the condensation energy and quasiparticle excitations as well as the Fermi liquid corrections.

\subsection{Effect of Fermi liquid corrections}

\begin{figure}[tb!]
\includegraphics[width=70mm]{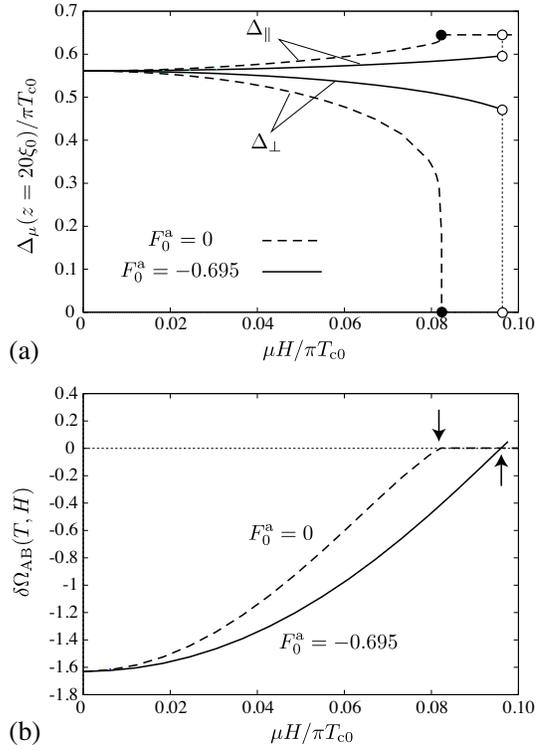}
\caption{Field-dependence of $\Delta _{\mu}(z)/\pi T_{\rm c0}$ at $ z\!=\! 20\xi_0$ (a) and of thermodynamic potential of the B-phase relative to the A-phase, $\delta \Omega _{\rm AB}$ (b), where $D \!=\!40 \xi _0$ and $T \!=\! 0.2T_{\rm c0}$. The solid (dashed) line denotes $\delta \Omega _{\rm AB}$ with (without) the Fermi liquid corrections and the arrows in (b) point the first- and second-order transition fields.}
\label{fig:omega}
\end{figure}

First, we emphasize that the thermodynamics is sensitive to the Fermi liquid corrections. Among their corrections, the $F^{\rm a}_0$ term associated with the local magnetization $M_{\mu}({\bm r})$ plays a crucial role. Figure~\ref{fig:omega}(a) shows the field dependence of $\Delta _{\mu}(z \!=\! D/2)$ of the squashed B-phase at $D \!=\! 40\xi _0 \!\approx\! 3.2\mu {\rm m}$. It is seen from Fig.~\ref{fig:omega}(a) that when the Fermi liquid corrections are absent, $\Delta _{\perp}(z \!=\! D/2)$ continuously vanishes at the critical field $\mu _{\rm n}H \!=\! 0.082\pi T_{\rm c0}$, where the second-order phase transition from B- to planar (or A-) phase occurs. Figure~\ref{fig:omega}(b) with the dashed line depicts the field-dependence of the thermodynamic potential introduced in Eq.~(\ref{eq:omega}), where $\delta \Omega _{\rm AB}(H,T)$ denotes the thermodynamic potential of the B-phase relative to the A-phase.

As seen in Fig.~\ref{fig:omega}(a) with solid lines, the Fermi liquid correction makes the field-dependence of $\Delta _{\mu}({\bm r})$ insensitive. This is because for a large $D$, the Fermi liquid correction associated with $M_{z}({\bm r})$ through $F^{\rm a}_0$ suppresses the effective magnetic Zeeman energy $H^{\rm eff}_{\mu}({\bm r}) \!<\! H$, except for the vicinity of the surface, as described in Eq.~(\ref{eq:Heff}). As a result of the suppression of the effective field, the B-phase survives even in the higher magnetic field so that the phase transition from the B- to A-phase turns to the first-order transition, as displayed in Fig.~\ref{fig:omega}(b) with the solid line. 

As $D$ decreases, however, the pair breaking effect at the specular surface gives rise to the distortion of the isotropic B-phase order parameter $\Delta _{\perp} \!<\! \Delta _{\parallel}$ even in low fields and the spin susceptibilities become comparable to the value in the normal $^3$He. In this case, the effective Zeeman energy $\mu _{\rm n}H^{\rm eff}$ is unchanged from that of the bare Zeeman field $\mu _{\rm n}H$ and the Fermi liquid correction does not alter the qualitative feature of the phase transition.

\subsection{Phase diagram}

\begin{figure}[tb!]
\includegraphics[width=70mm]{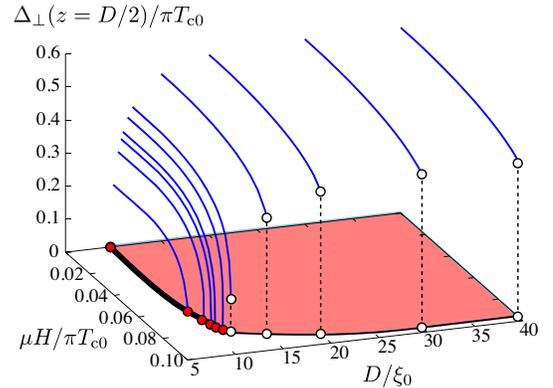}
\caption{(Color online) Field-dependence of $\Delta _{\perp}(z \!=\! D/2)$ for various values of the thickness $D$ at $T \!=\! 0.2\pi T_{\rm c0}$ (solid lines). The open (filled) circles denote the first-order (second-order) phase transition points. The bottom describes the phase boundary between the distorted B-phase (the shaded area) and the planar (or A-) phase, where the thick (thin) line corresponds to the first-order (second-order) line.}
\label{fig:phase02}
\end{figure}

Figure~\ref{fig:phase02} summarizes the field and thickness dependences of $\Delta _{\perp}(z \!=\! D/2)$ (solid lines) and the phase diagram (the bottom) at $T\!=\! 0.2 T_{\rm c0}$. In the region of the large thickness $D \!\gtrsim 11 \xi _0$, the phase boundary is the first-order phase transition $H _{\rm AB}$. As $D/ \xi_0$ increases, the first-order transition field $H _{\rm AB}$ slightly increases and reaches saturation $\mu _{\rm n}H^{\ast}_{\rm AB}/\pi T_{\rm c0} \!=\! 0.095$ in the thermodynamic limit $D\!\gg\! \xi _0$. Using the parameters $T_{\rm c0} \!=\! 1{\rm mK}$ and the gyromagnetic ratio of $^3$He nuclei $\gamma \!=\! 2\mu _{\rm n}$, the critical field is estimated as $H^{\ast}_{\rm AB} \!\approx\! 0.35 {\rm T}$, which is consistent to Ref.~\onlinecite{ashida} and experiments in Refs.~\onlinecite{kyynarainen} and \onlinecite{fisher}. 

As the thickness $D$ decreases the first-order transition turns to the second-order, where it is seen from Fig.~\ref{fig:phase02} that $\Delta _{\perp}(z \!=\! D/2)$ continuously touches zero at the critical field $H_{\rm AB}$. However, it is noted that the second-order phase transition may be proper to the weak coupling theory where the planar phase and A-phase are energetically degenerate. The finite contribution of an anisotropic interaction which makes A-phase more stable than the planar phase, such as the spin-fluctuation feedback effect, may change the phase boundary to the first-order transition even for small $D$'s. 

\begin{figure}[tb!]
\includegraphics[width=80mm]{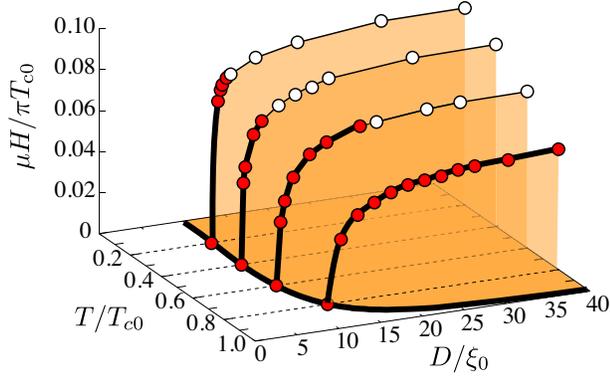}
\caption{(Color online) Superfluid phase diagram in the space spanned by temperature $T$, perpendicular magnetic field $H$, and thickness $D$. The definition of the open (filled) circles and thick (thin) lines are same as those in Fig.~\ref{fig:phase02}. The shaded area is occupied by the distorted B-phase and the other is covered by the planar (or A-) phase.}
\label{fig:phase}
\end{figure}

In Fig.~\ref{fig:phase}, we summarize the phase diagram in a three-dimensional space spanned by the temperature $T$, perpendicular magnetic field $H$, and thickness $D$. The first-order transition appears in the low temperature and large thickness region, while the high temperature and small thickness region involves the second-order phase transition. Note again that the thermodynamic limit of this phase diagram, corresponding to $D/\xi _0 \!\rightarrow \! \infty$, reproduces the well-known phase diagram in the bulk $^3$He which is composed of the first-order transition in low $T$'s and second-order line in high $T$'s.~\cite{ashida,kyynarainen}

The bottom line in Fig.~\ref{fig:phase} describes the A-B phase transition in the absence of a magnetic field. The whole line is found to be the second-order transition, which reproduces the earlier works done by Hara and Nagai in Ref.~\onlinecite{haraJLTP1988} and Vorontsov and Sauls in Ref.~\onlinecite{vorontsov}. However, Vorontsov and Sauls stated in Ref.~\onlinecite{vorontsovPRL} that the second-order phase boundary around $D \!\sim\! 10\xi _0$ is covered by the new quantum crystalline phase, the so-called stripe phase, in which the translational symmetry in the plane of the film is spontaneously broken. In this paper, for simplicity, we eliminate the possibility of the stripe phase from the phase diagram. Since the stability against a magnetic field is not trivial, the complete phase diagram which takes account of the stripe phase remains as a future problem.

\subsection{Spatially averaged spin susceptibilities}

\begin{figure}[tb!]
\includegraphics[width=80mm]{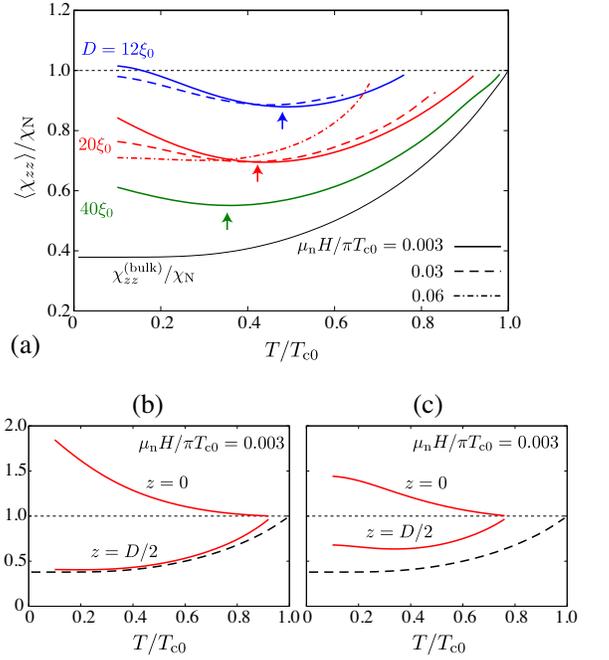}
\caption{(Color online) (a) $T$-dependence of the spatially averaged spin susceptibility $\langle \chi _{zz} \rangle$ at $\mu _{\rm n}H/\pi T_{\rm c0} \!=\! 0.003$ (solid lines), $0.03$ (dashed lines), and $0.06$ (dotted-dashed lines) for $D \!=\! 12 \xi _0$, $20\xi _0$, and $40\xi _0$. The arrows denote $T_{\rm min}$ in the lowest field (see the text). (b), (c) $T$-dependence of local spin susceptibilities $\chi _{zz}(z)$ at $z \!=\! 0$ and $D/2$; (b) $D\!=\! 20\xi_0$ and (c) $12\xi _0$ at $\mu _{\rm n}H \!=\! 0.003 \pi T_{\rm c0}$. The thin solid line in (a) and dashed line in (b) and (c) depicts the spin susceptibility $\chi^{({\rm bulk})}_{zz}$ in the bulk B-phase given in Eq.~(\ref{eq:chiB}).}
\label{fig:chiTavg}
\end{figure}

Figure \ref{fig:chiTavg}(a) shows the $T$-dependence of spin susceptibility $\langle \chi _{zz} \rangle$ averaged over the slab for $D/\xi _0 \!=\! 12$, $20$, and $40$, where $\langle \chi _{zz} \rangle$ is defined as
\beq
\langle \chi _{zz} \rangle \equiv \frac{1}{D}\int^{D}_0 \chi _{zz}(z) dz.
\eeq
For comparison, we plot the spin susceptibility in the bulk B-phase given with the Fermi liquid parameter $F^{\rm a}_0$ by
\beq
\chi^{({\rm bulk})}_{zz} = \frac{(1+F^{\rm a}_0)[2+Y(T)]}{3+F^{\rm a}_0[2+Y(T)]}\chi _{\rm N},
\label{eq:chiB}
\eeq
where $Y(T)$ is the Yosida function.~\cite{vollhardt} The nonlinear effect of the Zeeman magnetic field on $\chi^{({\rm bulk})}_{zz}$ was investigated by Fishman and Sauls in Ref.~\onlinecite{FS}.

According to the sum rule, the static spin susceptibility $\langle\chi _{zz}\rangle$ is obtained by integrating the absorptive part of the dynamical spin susceptibility over all the frequency.~\cite{Leggett} Hence, the temperature- and field-dependences are detectable through NMR experiments.~\cite{ahonen}

It is seen from Fig.~\ref{fig:chiTavg}(a) that in weak magnetic fields, {\it e.g.}, $\mu _{\rm n}H/\pi T_{\rm c0} \!=\! 0.003$, the spatially averaged spin susceptibility $\langle \chi _{zz} \rangle$ has a minimum value at a certain temperature, {\it e.g.}, $T \!\equiv\! T_{\rm min} \!\approx\! 0.4 T_{\rm c0}$ for $D\!=\! 20\xi_0$, and raises up with further decreasing the temperature. This is contrast to the $T$-dependence of the bulk B-phase associated with the Yosida function. The increase of $\langle\chi _{zz}\rangle$ in the low-$T$ regime is found to reflect the considerable contribution of the gapped surface bound state. To understand the non-monotonic behavior, we plot in Figs.~\ref{fig:chiTavg}(b) and \ref{fig:chiTavg}(c) the $T$-dependence of local spin susceptibilities $\chi _{zz}(z)$ at $z \!=\! 0$ and $D/2$, where we set $D\!=\! 20\xi_0$ in Fig.~\ref{fig:chiTavg}(b) and $12\xi _0$ in Fig.~\ref{fig:chiTavg}(c). In the case of $D \!=\! 20\xi _0$, the spin susceptibility at $z \!=\! 10 \xi _0$ traces the $T$-dependence of $\chi^{({\rm bulk})}_{zz}$ which corresponds to the bulk B-phase, and stays almost constant around $\chi _{zz}(z \!\sim\! D/2) \!\sim\! 0.4 \chi _{\rm N}$ in low temperatures within $T\!\lesssim\!0.4T_{\rm c0}$. As we have discussed in Sec.~IV B, a Zeeman magnetic field perpendicular to the surface opens the finite energy gap in the SABS, which gives rise to the enhancement of the spin susceptibility. In actual, it is seen from Fig.~\ref{fig:chiTavg}(b) that $\chi _{zz}(z)$ at the surface exceeds the Pauli susceptibility of the normal $^3$He in the low temperature regime, while it monotonically decreases as $T$ increases. Hence, the spin susceptibility $\langle \chi _{zz} \rangle$ averaged over the slab in the low $T$ region of Fig.~\ref{fig:chiTavg}(a) indicates the enhancement of local magnetization density at the surface, while the behavior in the high $T$ regime is dominated by the magnetization density in the central region of the system. 

As shown in Fig.~\ref{fig:chiTavg}(a), the qualitative feature on the $T$-dependence of $\langle\chi _{zz}\rangle$ is insensitive to the thickness $D$, except for the vicinity of the A-B transition $D_{\rm AB} \!\approx\!9.6\xi _0$ in which the magnetic response becomes indistinguishable from that in the normal $^3$He. It is demonstrated in Fig.~\ref{fig:chiTavg}(a) with the solid lines that the temperature $T_{\rm min}$, at which $\langle\chi _{zz}\rangle$ becomes minimum, lowers as $D$ increases, namely, the thermodynamic limit is approached. 

Then, let us look at the field-dependence of $\langle \chi _{zz} \rangle$, which is plotted in Fig.~\ref{fig:chiTavg}(a) with the dashed and dotted-dashed lines. In the bulk B-phase, as discussed in Refs.~\onlinecite{schopohl} and \onlinecite{FS}, the nonlinear effect of the Zeeman magnetic field enhances the spin susceptibility in the entire region of $T \!<\! T_{\rm c0}$ as $\chi _{zz}(H)-\chi _{zz}(0)\!\propto\! (\mu_{\rm n}H/\Delta _0(T))^2$.~\cite{FS} The $H$-dependence also appears in the high $T$ regime $T\!>\! T_{\rm min}$ of Fig.~\ref{fig:chiTavg}(a). In the regime of $T \!<\! T_{\rm min}$, however, the nonlinear effect of $H$ lowers the magnetization density at the surface as shown in Fig.~\ref{fig:mag}(b) and the resulting $\langle \chi _{zz}\rangle /\chi _{\rm N}$ is rather suppressed by increasing $H$. This implies that as the Zeeman magnetic field is ramped up, the characteristic temperature $T_{\rm min}$ gets lower and fades away at last. In summary, the field and temperature dependences of $\langle \chi _{zz}\rangle /\chi _{\rm N}$ may unveil the dispersion of the surface bound state in superfluid $^3$He-B.




\section{Concluding remarks}

Here, we have investigated the role of surface bound states on the thermodynamics and spin susceptibilities in $^3$He-B under a perpendicular magnetic field. First, within the Andreev approximation of the Bogoliubov-de Gennes equation, we have clarified the relation between the $SO(3)$ order parameter manifold and the surface bound state, where the direction of the Majorana Ising spin is clarified. We have also explicitly mentioned that the condition in which the surface bound state opens the maximum energy gap coincides with the condition that makes the magnetic field energy lower. 

Subsequently, we have revealed the thermodynamics and surface bound states in a restricted geometry. All the results are obtained with the quasiclassical Eilenberger theory which provides the closed set of selfconsistent equations reliable to the weak coupling regime of superfluid $^3$He. It turns out that the pair breaking effect and surface bound states play a crucial role on determining the phase diagram and spin susceptibilities. The Zeeman magnetic field perpendicular to the surface always opens a finite energy gap in the surface bound state. We have demonstrated that the gapped surface bound state gives rise to the positive contribution to the enhancement of the spin susceptibility at the surface, compared with that in the normal $^3$He. We have also emphasized the role of the Fermi liquid corrections in the phase diagram of a restricted geometry, which plays a critical role on determining the phase boundaries. 

We have also discussed the temperature- and field-dependences of the spatially averaged spin susceptibility. It is found that the local spin susceptibility in the central region of the sample obeys the ordinary Yosida function, while at the surface it considerably increases in the low temperature regime. Hence, the temperature-dependence of the spatially averaged spin susceptibility in the low temperature regime is dominated by the contribution of the surface bound state, leading to the nonmonotonic behavior. The characteristic temperature at which the spin susceptibility becomes minimum is sensitive to the thickness of the sample and monotonically decreases as the thickness increases. Furthermore, we have demonstrated that the nonlinear effect of the Zeeman magnetic field reduces the the spin susceptibility at the surface, resulting in the monotonic behavior on the temperature in the high field regime comparable with the A-B transition field. The spatially averaged spin susceptibility is detectable through NMR experiments. 

Finally, we would like to mention the issues of which we do not take account here: The effect of the surface boundary condition~\cite{thunneberg,buchholtzPRB1986,zhang,nagato98,nagatoJLTP98} and the possibility of the {\it stripe phase} with the spontaneous breaking of the translational symmetry.~\cite{vorontsovPRL} In the absence of a magnetic field, the surface density of states in the low energy region is considerably enhanced by the diffusive surface.~\cite{vorontsov,nagato98} The low-energy density of states filled in by the skew scattering of the quasiparticle at the rough surface might drastically change the temperature- and field-dependences of the spin susceptibility. Note that the specularity of the surface of $^3$He can be experimentally controlled by coating it with $^4$He layers.~\cite{murakawaPRL2009,murakawaJPSJ2011} Furthermore, the vicinity of the A-B phase transition around $D\!\sim\! 10\xi _0$ is occupied by the stripe phase,~\cite{vorontsovPRL} when the magnetic field is absent. However, the robustness against a Zeeman field is not trivial, which remains as a future problem. 

\section*{ACKNOWLEDGMENTS}

The author is grateful to M. Ichioka, K. Machida, and Y. Tsutsumi for fruitful discussions and comments. This work was supported by the "Topological Quantum Phenomena" (Grant No. 22103005) KAKENHI on Innovative Areas from MEXT of Japan.

\appendix

\section{Derivation of the dispersion in Eq.~(\ref{eq:sabs})}

Here, we describe the details about how to solve the Andreev equation (\ref{eq:andr}), 
\beq
\left[ - i \alpha v_{\rm F}\hat{k}_z\partial _z\underline{\tau}_z + \underline{V}
+ \underline{\Delta} ({\bm k}_{{\rm F},\alpha})\right] \tilde{\bm \varphi}_{\alpha}(z) = E\tilde{\bm \varphi}_{\alpha}(z),
\label{eq:andreev}
\eeq
where $\tilde{\bm \varphi}_{\pm} (z)$ describes the slowly varying part of quasiparticle wavefunction ${\bm \varphi}({\bm r})$, that is, ${\bm \varphi}({\bm r}) \!=\! \sum _{\alpha \!=\! \pm} C_{\alpha}\tilde{\bm \varphi}_{\alpha}(z)e^{i{\bm k}_{{\rm F},\alpha}\cdot{\bm r}}$ with ${\bm k}_{{\rm F},\alpha} \!=\! k_{\rm F}(\cos\!\phi _{\bm k}\sin\!\theta _{\bm k},\sin\!\phi _{\bm k}\sin\!\theta _{\bm k},\alpha\cos _{\bm k}\!\theta _{\bm k})$. The normalization condition is imposed on $\tilde{\bm \varphi}_{\alpha}(z)$ as 
\beq
\sum _{\alpha \!=\! \pm} \int \tilde{\bm \varphi}^{\dag}_{\alpha}(z) \tilde{\bm \varphi}_{\alpha}(z) dz = 1.
\label{eq:norm}
\eeq
The rigid boundary condition at $z \!=\! 0$, ${\bm \varphi}(x,y,z\!=\!0) \!=\! {\bm 0}$, leads to $C_{+} \!=\! -C_-$ and the continuity condition $\tilde{\varphi}_+(z \!=\! 0) \!=\! \tilde{\varphi}_-(z \!=\! 0)$.

Then, introducing $\underline{\mathcal{U}}(\hat{\bm n},\varphi)\!\equiv\! {\rm diag}[U(\hat{\bm n},\varphi),U^{\ast}(\hat{\bm n},\varphi)]$ and using the relation in Eq.~(\ref{eq:opsu2}), the BdG equation (\ref{eq:andreev}) within the Andreev approximation reduces to 
\beq
&& \hspace{-10mm}
\left[ - i\alpha v_{{\rm F},z} \partial _z\underline{\tau}_z + \underline{V}^{\prime} (\hat{\bm n},\varphi) 
+ \underline{\Delta}_0 ({\bm k}_{{\rm F},\alpha})\right] \nn \\
&&  \times \mathcal{U}^{\dag}(\hat{\bm n},\varphi){\bm \varphi}_{\alpha}(z) 
= E \mathcal{U}^{\dag}(\hat{\bm n},\varphi){\bm \varphi}_{\alpha}(z) ,
\label{eq:andreev1}
\eeq
where $v_{{\rm F},z} \!=\! v_{\rm F}\cos\theta _{\bm k}$. The $SU(2)$ matrix $\underline{\mathcal{U}}(\hat{\bm n},\varphi)$ in the Nambu representation rotates the Pauli matrices $\sigma _{\mu}$ and the Zeeman term $\underline{V}^{\prime} \!\equiv\!  \mathcal{U}^{\dag}(\hat{\bm n},\varphi)  \underline{V} \mathcal{U}(\hat{\bm n},\varphi)$ results in 
\beq
\underline{V}^{\prime} (\hat{\bm n},\varphi) = - \mu _{\rm n} H_{\mu}R_{\mu\nu}(\hat{\bm n},\varphi)\left(
\begin{array}{cc}
\sigma _{\nu} & 0 \\ 0 & - \sigma^{\ast}_{\nu}
\end{array}
\right).
\eeq
It is convenient to introduce the unitary matrix ${\mathcal{M}} \!\equiv \! (\sigma _x + \sigma _z ) e^{i\vartheta\sigma _z}/\sqrt{2}$ with $\vartheta \!=\! \frac{\phi _{\bm k}}{2} - \frac{\pi}{4}$. Then, the pair potential $\Delta _0 ({\bm k})\!=\!i\sigma _{\mu}\sigma_y\Delta_{\mu}\hat{k}_{\mu}$ in the B-phase rotates to $\Delta^{\prime}_0 ({\bm k}_{{\rm F},\alpha}) \!=\! M \Delta _0 ({\bm k}_{{\rm F},\pm})M^{\rm T}$, 
\beq
\Delta^{\prime}_0 ({\bm k}_{{\rm F},\alpha}) = \left[
\begin{array}{cc}
a (\phi _{\bm k},\theta _{\bm k}) & b (\phi _{\bm k},\theta _{\bm k}) \\ 
b (\phi _{\bm k},\theta _{\bm k}) & -a^{\ast}(\phi _{\bm k},\theta _{\bm k})
\end{array}
\right],
\label{eq:delta0}
\eeq
where $a (\phi _{\bm k},\theta _{\bm k}) \!=\! \Delta _z \cos\theta _{\bm k}\!+\! i(\Delta _x\cos^2\phi _{\bm k}+\Delta _y\sin^2\phi _{\bm k})\sin\theta _{\bm k}$ and $b (\phi _{\bm k},\theta _{\bm k}) \!=\!- (\Delta _x \!-\! \Delta _y)\sin\phi _{\bm k}\cos\phi _{\bm k}\sin\theta _{\bm k} $. For simplicity, let us assume that $\Delta _x \!=\! \Delta _y  \!=\! \Delta _z  \!=\! \Delta _0 \!\in \! \mathbb{R}$. This is valid for a weak field regime within $\mu _{\rm n}H \!\ll\! \Delta _0$, because the distortion induced by the magnetic field can be estimate in the thermodynamic limit as $\Delta _{\parallel}/\Delta _{\perp} \!=\! 1 - \mathcal{O}(\mu ^2_{\rm n}H^2/\Delta^2_0)$,~\cite{fishman} where $\Delta _{\parallel}$ ($\Delta _{\perp}$) represents the pair potential parallel (perpendicular) to an applied field. Within the assumption, Eq.~(\ref{eq:delta0}) reduces to $a(\phi _{\bm k},\theta _{\bm k}) \!=\! \alpha \Delta _0 e^{i\theta _{\bm k}}$ and $b (\phi _{\bm k},\theta _{\bm k}) \!=\! 0$. 

First, we solve the Andreev equation (\ref{eq:andreev}) in the absence of a magnetic field, $H\!=\! 0$. Then, the Andreev equation (\ref{eq:andreev1}) can be separated to two independent spin sectors as
\beq
{\bm \varphi}_{\alpha}(z) = a_f \left[
\begin{array}{c}  f^{(1)}_{\alpha} (z) \\ 0 \\f^{(2)}_{\alpha}(z) \\ 0 \end{array}
\right] + a_g  \left[
\begin{array}{c}   0 \\ g^{(1)}_{\alpha} (z) \\ 0 \\ g^{(2)}_{\alpha}(z) \end{array}
\right], 
\eeq
where $a_f$ and $a_g$ are the normalization constants. The equation for the wavefunctions $f^{(1,2)}_{\alpha}$ is obtained from Eq.~(\ref{eq:andreev1}) as
\beq
\mathcal{H}_0 ({\bm k}_{{\rm F},\alpha},z) 
\left[
\begin{array}{c}f^{(1)}_{\alpha} \\f^{(2)}_{\alpha} \end{array}
\right] = E_0 \left[
\begin{array}{c}f^{(1)}_{\alpha} \\f^{(2)}_{\alpha} \end{array}
\right],
\label{eq:oneD}
\eeq
where 
\beq
\mathcal{H}_0 ({\bm k}_{{\rm F},\alpha},z) 
= -i\alpha v_{{\rm F},z} \partial _z \sigma _z + \sigma _x \alpha \Delta _0 e^{-i\alpha\theta _{\bm k}\sigma _z}.
\eeq
Using the particle-hole symmetry, 
\beq
\sigma _x \mathcal{H}^{\ast}_0 ({\bm k},z) \sigma _x = - \mathcal{H}_0(-{\bm k},z) , 
\eeq
the positive energy states with the wavefunction $[f^{(1)}_{\alpha},f^{(2)}_{\alpha}]^{\rm T}$ and $E \!>\! 0$ is associated with the negative branch with $g^{(1,2)}_{\alpha} \!=\! \sigma _x [f^{(1)\ast}_{\alpha},f^{(2)\ast}_{\alpha}]^{\rm T}$ and $-E$.

The resulting equation (\ref{eq:oneD}) is equivalent to the one-dimensional Dirac equations with the mass domain wall. The index theorem~\cite{jackiw,tewari,stone} ensures the existence of the zero energy states when the {\it mass} term changes its sign. The bound state solution with $|E({\bm k}_{\parallel})| \!\le\! \Delta _0$ has the energy dispersion linear on the momentum ${\bm k}_{\parallel} \!=\! (k_x,k_y)$ as
\beq
E_0({\bm k}_{\parallel}) = \pm \frac{\Delta _0}{k_{\rm F}} |{\bm k}_{\parallel}|. 
\label{eq:E0}
\eeq 
This expression is independent of the orientation of $\hat{\bm n}$ and the angle $\varphi$. The corresponding wavefunctions for the quasiparticles bound at at $z\!=\! 0$ are given by
\beq
{\bm \varphi}^{(\pm)}_{0,{\bm k}_{\parallel}} ({\bm r}) = N_{\bm k}
e^{i{\bm k}_{\parallel}\cdot{\bm r}_{\parallel}}f(k_{\perp},z)\mathcal{U}(\hat{\bm n},\varphi)
{\bm \Phi}_{\pm}(\phi _{\bm k}), 
\label{eq:varphi}
\eeq
where $N_{\bm k}$ is the normalization constant estimated from Eq.~(\ref{eq:norm}). In Eq.~(\ref{eq:varphi}), we also set $f(k_{\perp},z)\!=\! \sin\left( k_{\perp}z\right)e^{-z/\xi}$ with $k_{\perp} \!\equiv\! \sqrt{k^2_{\rm F}-k^2_{\parallel}}$ and 
\beq
{\bm \Phi}_{\pm}(\phi _{\bm k}) \equiv e^{\pm i\frac{\phi _{\bm k}}{2}}\left[
e^{-i\frac{\phi _{\bm k}}{2}}
\left( 
\begin{array}{c}
1 \\ 0 \\ 0 \\ -i
\end{array}
\right)
\mp e^{i\frac{\phi _{\bm k}}{2}}\left( 
\begin{array}{c}
0 \\ i \\ 1 \\ 0
\end{array}
\right) \right].
\eeq
In Eq.~(\ref{eq:varphi}), ${\bm \varphi}^{(+)}_{0,{\bm k}_{\parallel}}$ corresponds to the positive energy solution and ${\bm \varphi}^{(-)}_{0,{\bm k}_{\parallel}}$ is the negative branch. The gapless spectrum of the SABS is protected by the nontrivial topological invariant defined in the bulk region of the B-phase in the absence of a magnetic field.~\cite{schnyder,qi}

Now let us consider the case of a finite magnetic field $H\!\neq\! 0$. 
\beq
{\bm \varphi}_{{\bm k}_{\parallel}}({\bm r}) = a_+{\bm \varphi}^{(+)}_{0,{\bm k}_{\parallel}}({\bm r}) + a_-{\bm \varphi}^{(-)}_{0,{\bm k}_{\parallel}}({\bm r}),
\label{eq:eigenfn}
\eeq
where the normalization condition for ${\bm \varphi}_{{\bm k}_{\parallel}}({\bm r})$ requires $|a_+|^2 + |a_-|^2\!=\! 1$. The coefficients $a_{\pm}$ and energy $E({\bm k}_{\parallel})$ are determined by solving the eigenvalue equation
\beq
\left(
\begin{array}{cc}
|E_0| & e^{-i\phi _{\bm k}} \gamma _z \\
e^{i\phi _{\bm k}} \gamma _z & -|E_0|
\end{array}
\right)\left( 
\begin{array}{c} a_+ \\ a_- \end{array}
\right) = E \left( 
\begin{array}{c} a_+ \\ a_- \end{array}
\right),
\label{eq:eigen}
\eeq
where $\gamma _z \equiv \mu _{\rm n}H_{\mu}R_{\mu z}(\hat{\bm n},\varphi)$ denotes the gap of the surface cone. From Eq.~(\ref{eq:eigen}), the dispersion of the SABS is given as
\beq
E({\bm k}_{\parallel}) = \pm \sqrt{ \left| E_0({\bm k}_{\parallel})\right|^2 + \left|\mu _{\rm n}H\hat{\ell}_z(\hat{\bm n}, \varphi)\right|^2},
\label{eq:sabs_app}
\eeq
and the wave functions are obtained from Eq.~(\ref{eq:eigenfn}) with $a_{\pm}({\bm k}_{\parallel}) \!=\! \sqrt{\frac{1}{2}(1\pm |\frac{E_0({\bm k}_{\parallel})}{E({\bm k}_{\parallel})}|)}$. Here, we introduce the $\hat{\bm \ell}$-vector in Eq.~(\ref{eq:sabs_app}), the definition~\cite{volovik2010,TM2012} of which is
\beq
\hat{\ell}_{\mu} (\hat{\bm n}, \varphi) \equiv \frac{H_{\nu}}{H} R_{\nu \mu} (\hat{\bm n}, \varphi).
\eeq

\section{Boundary conditions and numerical procedures}

The quasiclassical Green's function $\underline{g}$ is parameterized with $2\!\times\! 2$ matrices $a \!\equiv\! a(\hat{\bm k},{\bm r}; i\omega _n)$ and $b \!\equiv\! b(\hat{\bm k},{\bm r}; i\omega _n)$ as
\beq
\underline{g}(\hat{\bm k},{\bm r}; i\omega _n) = -i\pi \underline{N}
\left(
\begin{array}{cc}
\sigma _0 + ab & 2a \\ -2b & -\sigma _0 + ba
\end{array}
\right),
\eeq
where 
\beq
\underline{N} = \left[ 
\begin{array}{cc}
(\sigma _0 - ab)^{-1} & 0 \\ 0 & (-\sigma _0 + ba)^{-1}
\end{array}
\right].
\eeq
This Ricatti parametrization automatically satisfies the normalization condition of $\underline{g}$ and simplifies the Eilenberger equation (\ref{eq:eilen}) where the equations which governs $a$ and $b$ are separated to each other. The resulting equations, called the matrix Ricatti equations,~\cite{nagato93,eschrig99,eschrig00} are given by 
\begin{subequations}
\label{eq:ab}
\beq
i{\bm v}_{\rm F}(\hat{\bm k}) \cdot {\bm \nabla} a + 2i\omega _n a + \Delta - a \Delta^{\dag} a + a \tilde{\nu}^{\prime} - \tilde{\nu} a = 0,
\eeq
\beq
i{\bm v}_{\rm F}(\hat{\bm k}) \cdot {\bm \nabla} b - 2i\omega _n b + \Delta^{\dag} - b \Delta b + b \tilde{\nu} - \tilde{\nu}^{\prime} b = 0,
\eeq
\end{subequations}
where we set $\Delta^{\dag} \!\equiv\! \Delta^{\dag}(-\hat{\bm k},{\bm r})$ and $\tilde{\nu}\equiv\tilde{\nu}(\hat{\bm k},{\bm r})$ is composed of the Fermi liquid correction and the Zeeman energy
\begin{subequations}
\beq
\tilde{\nu} \equiv \nu _0 \sigma _0 + \nu _{\mu}\sigma _{\mu} - \frac{1}{1+F^{\rm a}_0}\mu _{\rm n}H_{\mu}\sigma _{\mu},
\eeq
\beq
\tilde{\nu}^{\prime} \equiv \nu^{\prime}_0 \sigma _0 + \nu^{\prime}_{\mu}\sigma _{\mu} - \frac{1}{1+F^{\rm a}_0}\mu _{\rm n}H_{\mu}\sigma^{\ast}_{\mu}.
\eeq
\end{subequations}
It is worth mentioning that the Ricatti amplitudes $a$ and $b$ have the following symmetry $a(\hat{\bm k},{\bm r};i\omega _n) \!=\! b^{\ast}(-\hat{\bm k},{\bm r};i\omega _n)$, which implies that the quasiclassical Green's functions obey 
\begin{subequations}
\beq
g_j(\hat{\bm k},{\bm r};i\omega _n) \!=\! \left[g^{\dag}_j(-\hat{\bm k},{\bm r};i\omega _n)\right]^{\ast}, 
\eeq
\beq
f_{\mu}(\hat{\bm k},{\bm r};i\omega _n) \!=\! \left[f^{\dag}_{\mu}(-\hat{\bm k},{\bm r};i\omega _n)\right]^{\ast}.
\eeq
\end{subequations}
where $j\!=\! 0, x, y, z$. 

In this work, we consider superfluid $^3$He sandwiched by two specular surfaces, as displayed in Fig.~\ref{fig:system}. Assuming spatial uniformity in the plane parallel to the surfaces, the resulting Ricatti equations (\ref{eq:ab}) reduce to one-dimensional ordinary differential equations along the $\hat{\bm z}$ axis, which are numerically stable and requires an initial value of $a$ and $b$. For a quasiparticle momentum $\hat{\bm k}$ on three-dimensional Fermi sphere with ${\bm v}_{\rm F}(\hat{\bm k}) \!=\! v_{\rm F}\hat{\bm k}$, we solve the Ricatti equations by numerically integrating along the classical forward (backward) trajectories for $a$ ($b$) with an arbitrary initial value. The numerical integration of one-dimensional Ricatti equations (\ref{eq:ab}) is performed with the fourth-order Runge-Kutta method from an arbitrary point of $z$. An arbitrary initial value of $a$ and $b$ converges after multiple reflections on the specular surfaces situated at $z \!=\! 0$ and $D$. The specular surface requires the matching of two propagators $a(\hat{\bm k},z;i\omega _n)$ and $a(\underline{\hat{\bm k}},z;i\omega _n)$, that is,
\beq
a(\hat{\bm k},z;i\omega _n) = a(\underline{\hat{\bm k}},z;i\omega _n), \hspace{3mm} \mbox{for $z = 0$ and $D$},
\label{eq:bc}
\eeq
where $\underline{\hat{\bm k}} \!=\! (\cos\phi _{\bm k}\sin\theta _{\bm k}, \sin\phi _{\bm k}\sin\theta _{\bm k},-\cos\theta _{\bm k})$. The boundary condition on $b$ is given in the same way as Eq.~(\ref{eq:bc}).

\section{$SO(2)$ rotational invariance of the quasiclassical Green's functions}

Here, we clarify the symmetric property of the quasiclassical Green's functions. First of all, in the absence of a Zeeman magnetic field $\underline{v} \!=\! 0$, the order parameter in the B-phase is isotropic in the sense of $\Delta _{\mu} \!\equiv\! \Delta _0$. Here, we introduce the simultaneous rotation in the orbital space $R^{({\bm L})}_{\mu\nu} \!=\! O_{\mu\nu}$ and spin space $R^{({\bm S})}_{\mu\nu} \!=\! (ROR^{-1})_{\mu\nu}$ where $(R)_{\mu\nu} \!\equiv\! R_{\mu\nu}$.~\cite{vollhardt} Using the rotation matrices, the momentum $\hat{\bm k}$ and the Pauli matrices $\sigma _{\mu}$ (or equivalently the ${\bm d}$-vector) behave as the three-dimensional vectors, which are transformed to $\hat{k}_{\mu} \!\mapsto\! R^{({\bm L})}_{\mu\nu}\hat{k}_{\nu}$ and $\hat{\sigma}_{\mu} \!\mapsto\! R^{({\bm S})}_{\mu\nu}\hat{\sigma}_{\nu}$. Now let $U_{\bm S}$ be an $SU(2)$ representation of the $SO(3)$ rotation matrix $R^{({\bm S})}_{\mu\nu}$. Then, the isotropic B-phase order parameter with $\Delta _{\mu} \!\equiv\! \Delta _0$ is invariant under the joint rotation of spin and orbital spaces, $SO(3)_{{\bm L}+{\bm S}}$, 
\beq
\Delta (\hat{\bm k},{\bm r}) = i\sigma _{\mu}\sigma _y \Delta _0 R_{\mu\nu}\hat{k}_{\nu}
= U_{\bm S}\Delta (R^{({\bm L})}\hat{\bm k},{\bm r})U_{\bm S}^{\rm T}.
\eeq

However, a magnetic field and surface boundary condition reduces the joint rotational symmetry $SO(3)_{{\bm L}+{\bm S}}$. Since we consider the situation where a magnetic field is applied along the surface normal (${\bm H} \!\perp\! \hat{\bm z}$) and the dipole interaction is absent, it is natural to suppose that the components of the B-phase pair amplitudes still remains isotropic about the $\hat{z}$-axis, $\Delta _x \!=\! \Delta _y \!\equiv\! \Delta _{\parallel}$ and $\Delta _z \!\equiv\! \Delta _{\perp} \!\neq\! \Delta _{\parallel}$. Let $O^{(2)}$ be a two-dimensional rotation matrix around the $z$-axis and $U^{(2)}_{\bm S}$ be an $SU(2)$ representation of the $SO(2)$ rotation matrix $(RO^{(2)}R^{-1})_{\mu\nu}$.  Then, it turns out that the squashed B-phase order parameter is invariant under the two-dimensional rotation of spin and orbital spaces around the $z$-axis, $SO(2)_{L_z + S_z}$,
\beq
\Delta (\hat{\bm k},{\bm r}) =  U^{(2)}_{\bm S}\Delta (O^{(2)}\hat{\bm k},{\bm r})U^{(2){\rm T}}_{\bm S}.
\label{eq:so2}
\eeq

Now let us apply the $SO(2)_{L_z+S_z}$ rotation to the quasiclassical Green's functions as $\underline{\mathcal{U}_2} \underline{g}(O^{(2)}\hat{\bm k},{\bm r};i\omega _n)\underline{\mathcal{U}_2}^{\dag} \!\equiv\! \underline{\tilde{g}}(O^{(2)}\hat{\bm k},{\bm r};i\omega _n)$, where we introduce $\underline{\mathcal{U}_2} \!\equiv\! {\rm diag}[U^{(2)}_{\bm S}, U^{(2)\ast}_{\bm S}]$ in the Nambu representation. Then, the Pauli matrices $\sigma _{\mu}$ and the momentum $\hat{k}_{\mu}$ are transformed to $\tilde{\sigma}_{\mu} \!=\! (RO^{(2)}R^{-1})_{\mu\nu}\sigma _{\nu}$ and $\hat{\tilde{k}}_{\mu} \!=\! O^{(2)}_{\mu\nu}\hat{k}_{\nu}$. The quasiclassical self-energy $\underline{S}$ in Eq.~(\ref{eq:eilen}) is transformed to $\underline{\mathcal{U}_2} \underline{\mathcal{S}}(O^{(2)}\hat{\bm k},{\bm r})\underline{\mathcal{U}_2}^{\dag} \!\equiv\!\tilde{\mathcal{S}}(O^{(2)}\hat{\bm k},z)$ in the same way. The magnetic Zeeman term in Eq.~(\ref{eq:eilen}) with ${\bm H} \!=\! H \hat{\bm z}$ is invariant under the $SO(2)_{L_z+S_z}$ rotation. To this end, the Eilenberger equation (\ref{eq:eilen}) under the $SO(2)_{L_z+S_z}$ rotation reduces to 
\beq
&&\hspace{-15mm}
\left[ 
i\omega _n \underline{\tau}_z - \tilde{\mathcal{S}}(O^{(2)}\hat{\bm k},z) - \underline{v}, 
\underline{\tilde{g}}(O^{(2)}\hat{\bm k},z; i\omega _n) 
\right] \nn \\
&& \hspace{10mm} + iv_{\rm F}\hat{k}_z\partial _z \underline{\tilde{g}}(O^{(2)}\hat{\bm k},z; i\omega _n)  = \underline{0}.
\eeq

Under the $SO(2)_{L_z+S_z}$ rotation, the quasiclassical self-energy matrix $\mathcal{S}(\hat{\bm k},z)$ mapped to 
\beq
\tilde{\mathcal{S}}(O^{(2)}_{\mu\nu}\hat{k}_{\nu},z) = \left[
\begin{array}{cc}
\tilde{\nu}(O^{(2)}_{\mu\nu}\hat{k}_{\nu},z) & \Delta (\hat{\bm k},z) \\ \Delta^{\dag}(-\hat{\bm k},z) & \tilde{\nu}^{\dag}(-O^{(2)}_{\mu\nu}\hat{k}_{\nu},z)
\end{array}
\right],
\eeq
where we use Eq.~(\ref{eq:so2}) and the diagonal part is 
\beq
\tilde{\nu}(O^{(2)}\hat{\bm k},z) = 
\nu _0 (O^{(2)}\hat{\bm k},z) \sigma _0 
+ \nu _{\mu} (O^{(2)}\hat{\bm k},z) \tilde{\sigma}_{\mu}.
\label{eq:nu2}
\eeq
As described in Eq.~(\ref{eq:nu}), the terms $\nu _0$ and $\nu _{\mu}$ are expanded in terms of the Legendre polynomials $P_{\ell}$. Among the possible contributions, we suppose in this paper that only the $\ell \!=\! 0$ and $1$ channels play a crucial role on thermodynamics and surface bound states, and the contributions with the higher $\ell$'s are eliminated. For $^3$He-B, one finds $\nu _0 \!=\! 0$ because of the absence of the mass flow. Hence, Eq.~(\ref{eq:nu2}) reduces to 
\beq
\tilde{\nu}(O^{(2)}\hat{\bm k},z) &=& -\mu _{\rm n} A^{({\rm a})}_0 \left[ H_{\mu} - M_{\mu}(z)\right] \tilde{\sigma}_{\mu} \nn \\
&& + \frac{A^{({\rm a})}_1}{2v_{\rm F}}J_{\mu\nu}(z) \tilde{\sigma}_{\mu}O^{(2)}_{\nu\eta}\hat{k}_{\eta}.
\eeq
Now, we suppose that the quasiclassical self-energy $\nu$ is invariant under the $SO(2)_{L_z+S_z}$ rotation in $^3$He-B with a perpendicular magnetic field,
\beq
\tilde{\nu}(O^{(2)}\hat{\bm k},z) =\nu(\hat{\bm k},z).
\eeq
This requires that the magnetization density $M_{\mu}$ and the spin current $J_{\mu\nu}$ must satisfy the following conditions:
\beq
M_x (z) = M_y(z) = 0, \hspace{3mm} J_{xy}(z) = - J_{yx} (z).
\label{eq:so2cond}
\eeq
With the selfconsistent calculation of the quasiclassical Eilenberger equations, we confirmed that the B-phase under a perpendicular magnetic field always satisfies the conditions in Eq.~(\ref{eq:so2cond}).


To this end, the Eilenberger equation within the $SO(2)_{L_z+S_z}$ symmetry is written as
\beq
&&\hspace{-15mm}
\left[ 
i\omega _n \underline{\tau}_z - \mathcal{S}(\hat{\bm k},z) - \underline{v}, 
\underline{\tilde{g}}(O^{(2)}\hat{\bm k},z; i\omega _n) 
\right] \nn \\
&& \hspace{10mm} + iv_{\rm F}\hat{k}_z\partial _z \underline{\tilde{g}}(O^{(2)}\hat{\bm k},z; i\omega _n)  = \underline{0},
\label{eq:eilenso2}
\eeq
which gives the equation for the quasiclassical Green's function at a point $O^{(2)}\hat{\bm k}$ of the three-dimensional Fermi sphere. Equation (\ref{eq:eilenso2}) is also equivalent to the equation for $\underline{g}(\hat{\bm k},z; i\omega _n)$. Namely, there is a one-to-one correspondence of the quasiclassical Green's functions between two points $\hat{\bm k}$ and $O\hat{\bm k}$ and the quasiclassical Green's function at an arbitrary point of the Fermi sphere, $O^{(2)}\hat{\bm k}$, is obtained from $\underline{g}(\hat{\bm k},z; i\omega _n)$ as
\beq
\underline{g}(O^{(2)}\hat{\bm k},z; i\omega _n) = \underline{\mathcal{U}}^{\dag}_2 \underline{g}(\hat{\bm k},z; i\omega _n)\underline{\mathcal{U}}_2. 
\label{eq:Gso2}
\eeq
This relation through the $SO(2)$ rotation is useful for shorting the computation time of the selfconsistent calculation. Once we calculate $\underline{g}(\hat{\bm k},z; i\omega _n)$ along the path (i) displayed in Fig.~\ref{fig:sphere}, the Green's function $\underline{g}$ for all $\hat{\bm k}$ is given by the symmetric relation in (\ref{eq:Gso2}) with $\underline{g}(\hat{\bm k},z; i\omega _n)$. 


\end{document}